\documentclass[%
 aip,
 amsmath,amssymb,
 reprint,%
]{revtex4-1}

\usepackage{graphicx}
\usepackage{dcolumn}
\usepackage{bm}

\usepackage[utf8]{inputenc}
\usepackage[T1]{fontenc}
\usepackage{mathptmx}
\usepackage{color}
\usepackage{etoolbox}

\makeatletter
\def\@email#1#2{%
 \endgroup
 \patchcmd{\titleblock@produce}
  {\frontmatter@RRAPformat}
  {\frontmatter@RRAPformat{\produce@RRAP{*#1\href{mailto:#2}{#2}}}\frontmatter@RRAPformat}
  {}{}
}%
\makeatother

\begin{document}

\title[]{On the selective formation of cubic tetrastack crystals 
from tetravalent patchy particles}

\author{{\L}ukasz Baran}
\email{lukasz.baran@mail.umcs.pl}
\author{Dariusz Tarasewicz}
\affiliation
{Department of Theoretical Chemistry, Institute of Chemical Sciences, Faculty of Chemistry, Maria-Curie-Sklodowska University in Lublin, Pl. M Curie-Sklodowskiej 3, 20-031 Lublin, Poland}
\author{Daniel M. Kami{\'n}ski}
\affiliation
{Department of Organic and Crystalochemistry, Institute of Chemical Sciences, Faculty of Chemistry, Maria-Curie-Sklodowska University in Lublin, Pl. M Curie-Sklodowskiej 3, 20-031 Lublin, Poland}
\author{Andrzej Patrykiejew}
\affiliation
{Department of Theoretical Chemistry, Institute of Chemical Sciences, Faculty of Chemistry, Maria-Curie-Sklodowska University in Lublin, Pl. M Curie-Sklodowskiej 3, 20-031 Lublin, Poland}
\author{Wojciech R{\.z}ysko}
\affiliation
{Department of Theoretical Chemistry, Institute of Chemical Sciences, Faculty of Chemistry, Maria-Curie-Sklodowska University in Lublin, Pl. M Curie-Sklodowskiej 3, 20-031 Lublin, Poland}

\date{\today}



\begin{abstract}

    Achieving the formation of target open crystalline lattices from colloidal particles is
    of paramount importance for their potential application in photonics. Examples of such desired structures
    are the diamond, tetrastack, and pyrochlore lattices. Herein, we demonstrate that the 
    self-assembly of tetravalent patchy particles results in the selective formation of the cubic tetrastack crystals,
    both in the bulk and in the systems subjected to external fields exerted by the solid substrate. 
    It is demonstrated that the presence of external field allows for the formation of  well-defined 
    single crystals with a low density of defects. Moreover, depending on the strength of the applied external field, 
    the mechanism of epitaxial growth changes. For weakly attractive external fields, the crystallization occurs in a similar manner as 
    in the bulk, since the fluid does not wet the substrate. Nonetheless, the formed crystal is considerably better ordered than the 
    crystals formed in the bulk, since the surface induces the ordering in the first layer.
    On the other hand, it is demonstrated that the formation of well ordered cubic tetrastack crystals is considerably enhanced by the increase of the
    external field strength and the formation of the thick crystalline film occurs via 
    a series of layering transitions.

\end{abstract}

\maketitle

\section{Introduction} 
\label{INTRODUCTION}

It is well known that the self-assembly of colloidal particles is primarily governed by their shape and surface properties. Isotropic spherical colloids
usually crystallize into simple face-centered cubic (fcc) and hexagonal close-packed (hcp) crystals, depending on the system properties and thermodynamic
conditions\cite{CR0}. This behavior results from the similarity of such colloids to hard-sphere systems\cite{CR1}. 
Since the spherical isotropic particles form only a very limited number of ordered structures, 
much of the recent interest focuses on anisotropic colloidal particles, which can be used as building blocks for new functional materials.    
The particles characterized by different shapes can now be obtained using many methods\cite{A1,A2,A3,A4,A5,A8}, and are known to assemble into
a variety of complex structures\cite{A7}. However, there is a class of spherical colloidal particles with anisotropic interactions, which 
comprises the so-called patchy colloids, with chemically and/or physically modified surfaces\cite{A7,jan1,jan2,S2,S3}.   
Already the simplest patchy particles with only one attractive patch, 
known as the Janus particles, exhibit the surface chemical anisotropy, which can be tuned by the appropriate
functionalization\cite{jan2,jan3}.  Both, the size and the chemical nature
of attractive patch, influence self-assembly of Janus particles, and lead to the formation of various ordered structures, 
in two- and three-dimensional systems\cite{jan3,size1,jan5,jan7,kagome}. 
In the region of low and moderate densities, 
the  formation of micelles, vesicles, and worm-like clusters have been observed\cite{ord1,ord2,ord3}. 
It has also been shown that Janus particles form crystals of different structure and density\cite{ord3}.

The situation becomes more complex when the number of attractive patches is larger\cite{S3,S4}. For example, the dissipative particle dynamics simulation of quasi two-dimensional  
triblock nanoparticles, with two attractive patches located at the poles of each particle, have shown the formation of two different ordered structures\cite{Kag1} of different densities. 
In the ordered phase of lower density,  the particles are arranged on the Kagom{\'e} lattice, while the high density phase shows the close-packed hexagonal ordering. 
In three-dimensional systems, the number of possible ordered structures formed by triblock particles is larger, and 
depends also on the size of attractive patches\cite{3d-tri-1,3d-tri-2,3d-tri-3,3d-tri-4}. 
The size and shape of attractive patches determines the valence, and hence the structure of crystal lattices\cite{3d-tri-3, ph1, ph2}.

A particularly interesting class of ordered structures that can be obtained via self assembly of patchy colloids are the open lattice photonic band gap crystals\cite{ph1,ph2,ph3}.
Such materials may find numerous applications in the devices allowing to control the propagation of light\cite{light}.
Among the most promising are the crystals of diamond\cite{diam1,diam2}, tetrastack\cite{3d-tri-4, tetrastack1, A6}, and pyrochlore\cite{pyr1,pyr2, liu2023} structures. 
The basic problem with such open crystals is that their synthesis is 
often hampered by low free energy differences between different polymorphic forms. The open structures resulting from the self-assembly of patchy particles are
stabilized by larger rotational and vibrational entropy, than in close-packed crystals\cite{ent1,ent2}. The synthesis of three-dimensional colloidal crystals with open lattices
is considerably hampered since it usually leads to large numbers of stacking faults and grain boundaries.
A possible way to force the formation of colloidal crystals with a desired lattice structure, and possibly low defect density, is the use of external fields or templates.
Trau et al.\cite{tau} developed an electrohydrodynamic method enabling the assembly of multilayer colloidal crystals on electrode surfaces. Another possibility is the 
use of patterned substrates as templates
for the assembly of three-dimensional colloidal crystals\cite{temp1,temp2,temp3,MC-temp}. The main advantage of epitaxial assembly is the formation of well ordered crystals with very low defect densities.
The problem with such approaches is, however,  the necessity of using
the templates being strictly commensurate with the assumed structure of colloidal crystals. Such methods have been primarily used to obtain open lattice colloidal
crystals by sedimentation of binary mixtures of colloidal particles characterized by different sizes, and a subsequent removal of one of the components. 
It has recently been shown by our group \cite{LB1} that the presence of a uniform external field, exerted by the solid substrate, allows to obtain a cubic diamond in systems
of tetrahedral patchy particles. A key element of the successful growth of cubic diamond crystals is the structure of the first adsorbed layer, being commensurate with the [110]
face of the cubic diamond. It has also been demonstrated that the subsequent removal of the external field does not affect the generated structure paving the way for further post-synthetic treatments.

In this paper, we use molecular dynamics simulations to investigate the formation of other open lattice crystals by adsorption of patchy particles on a structureless attractive substrate.
The only role of the substrate is to provide an external field leading to the adsorption of particles, but the emerging structure of adsorbed films is entirely 
determined by the assumed structure of patchy particles, and inter-particle interactions governing their self-assembly. It is well known that the interaction 
between colloidal particles is rather short-ranged\cite{Grzybowski}, but the external fields which determine the formation of ordered structures are usually of longer range. 
In particular, it is the gravity that leads to self-assembly resulting from the sedimentation of colloidal particles immersed in suspensions. 

In this work, it is shown that the self-assembly of tetravalent model patchy particles leads to the selective formation of cubic tetrastack (CT) structure in the bulk as well as in the external field.
However, the formation of well-ordered crystals in the bulk is hampered by the presence of numerous defects and grain boundaries between nucleating crystals. On the 
other hand, when ``the disordered fluid'' of patchy particles is placed in contact with the solid substrate, it forms quite well-ordered crystals, via a sequence of 
layering transitions, which occur upon the decrease in temperature.


\section{Methods}

\subsection{Patchy particle model}
\label{sec:patch}

\begin{figure}[h!]
    \centering
    \includegraphics[width=0.8\linewidth]{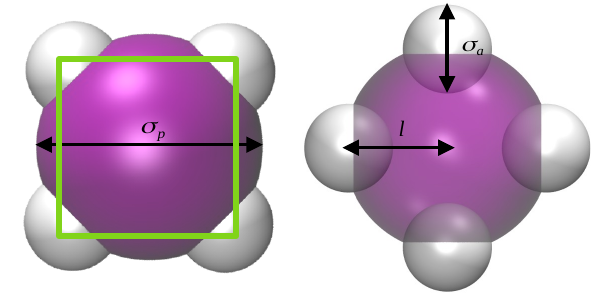}
    \caption{Schematic representation of a model of patchy particles and its parameters. The central spherical core and active sites are shown by purple and white spheres, respectively. The green square depicts the planar surface formed by the centers of active sites. Scale is not preserved for better clarity of the sketch.}
    \label{fig:model}
\end{figure}

The patchy particles model used in our simulations is identical to that used in our previous papers \cite{LB1}
with the only difference being the patch's arrangement.
Each particle is comprised of a spherical unit of a diameter $\sigma_p$ on which surface
four patches are displaced, each of a diameter $\sigma_a$ that are embedded to the certain extent. 
The latter can be manipulated by the parameter $l$. A schematic representation of model parameters
is shown in Figure~\ref{fig:model}. 
The active sites are arranged on a square on the surface intersecting the central
plane of the sphere. 
{\color{black} All the entities in the} patchy particles interacted via truncated and shifted Lennard Jones (12,6) potential.
$\sigma_p=\sigma$ and $\varepsilon_{pp}=\varepsilon$ were defined to be units of 
length and energy, respectively. In the current case, we have chosen the set of parameters
to be equal $\sigma_{a}=0.2\sigma$, $\sigma_p=1.0\sigma$ and $l=0.45\sigma$.
Parameter $l=0.45\sigma$ ensures that every active site can form three bonds
resulting in twelve neighbors in total.

The association energy was fixed to $\varepsilon_{aa}=5.0\varepsilon$ whereas $\varepsilon_{pp}=\varepsilon_{ap}=
1.0\varepsilon$. The range of the interactions was set to {\color{black}$r_{cut,aa}=2.0\sigma_{aa}$} and
the remaining were $r_{cut,ij}=1.0\sigma_{cut,ij}$ with ($ij,=pp,ap$). 
Choosing this set of parameters ensures that the only attraction in the system is 
due to the association energy, while the remaining interactions are soft-repulsive. 

To maintain the rigidity and the geometry of patchy particles we used 
harmonic spring potentials for bonds and bond angles. The corresponding constants
were equal to $k_b=1000\varepsilon/\sigma^2$ and $k_\theta=1000\varepsilon/\mathrm{rad^2}$
for bonds and bond angles, respectively, ensuring negligible fluctuations.

To study the surface-directed self-assembly of patchy particles, 
we introduced the structureless external wall    
modeled by the Lennard-Jones (9,3) potential, defined as follows:

\begin{equation}
 U^{\mathrm{ext}}(z)=\varepsilon_{wc} \left [\frac{2}{15} \left (\frac{\sigma}{z} \right)^9 - \left (\frac{\sigma}{z} \right)^3 \right ] 
 ~~~~~~~{\rm if~} z\leq5\sigma
\end{equation}

\noindent where $\varepsilon_{wc}$ indicates the depth of potential well for interactions.

The potential acted solely on the core of the patchy particle 
which should result in the introduction of isotropy on how the
patchy particles will be arranged with respect to the wall. 
The following results will be described in terms of the 
change in the ratio $\xi$ of attraction strengths 
defined as $\xi=\varepsilon_{wc}/\varepsilon_{aa}$.

The table~\ref{tab:parameters} gives a full overview of the system parameters.

\begin{table}[h!]
    \centering
    \begin{tabular}{c|c|c}
    \hline
    \hline
        parameter & symbol  & value \\
        \hline
        core diameter & $\sigma_p$ & $1.0\sigma$ \\
        active site diameter & $\sigma_a$ & $0.2\sigma$ \\
        embedding distance & $l$  & $0.45\sigma$  \\
        association energy & $\varepsilon_{aa}$  & $5.0\varepsilon$  \\
        remaining energies & $\varepsilon_{ij}$  & $1.0\varepsilon$  \\
        association cutoff & $r_{cut,aa}$  & $2.0\sigma_{aa}$  \\
        remaining cutoffs & $r_{cut,ij}$  & $1.0\sigma_{ij}$  \\
        bond harmonic constant & $k_b$  & $1000\varepsilon/\sigma^2$  \\
        bond angle harmonic constant & $k_\theta$  & $1000\varepsilon/\mathrm{rad}^2$  \\
        external field energy & $\varepsilon_{wc}$ & $2.0\varepsilon-8.0\varepsilon$ \\
        \hline
        \hline
    \end{tabular}
    \caption{Parameters of the model. In the above $ij=a,p$.}
    \label{tab:parameters}
\end{table}

\subsection{3D order parameter}
\label{sec:3d-ord}
To detect the formed three-dimensional tetrastack
crystalline network, we employed a routine based on
the calculation of Steinhardt order parameter \cite{steinhardt} given by:

\begin{equation}
 q_{l}=\sqrt{\frac{4\pi}{2l+1} \sum_{m=-l}^{m=l} \left| q_{lm} \right|^2}
 \label{eq:stein}
\end{equation}

\noindent with

\begin{equation}
 q_{lm}=\frac{1}{N_b(i)} \sum_{j \in N_b(i)} 
 Y_{lm}(\theta_{ij}, \phi_{ij}) 
 \label{eq:stein1}
\end{equation}

\noindent where $Y_{lm}$ are spherical harmonics. For a given sphere $i$, we choose a set of its 
nearest neighbors, $N_b(i)$. We define that any two spherical particles are connected by a bond if they are neighbors, that is, if $ j \in N_b(i)$. 
Each vector $\mathbf{n}_{ij}$ is characterized by its angles in spherical coordinates $\theta_{ij}$ and $\phi_{ij}$ on the unit sphere, evaluated
between the bond and an arbitrary but fixed reference frame.
The set of all bond vectors is called the bond network.

The ideal tetrastack network is comprised of alternating hexagonal and Kagom\'e layers. 
In the crystal lattice, we identify two types of local environments where every particle has either
10 or 12 nearest neighbors (NN). We found that the set of $q_4$ and $q_6$ are the best for the discrimination.
For ideal lattice for $NN=12$, Steinhardt parameters are equal to $q_4=0.206$ and $q_6=0.558$
and we allow for the $10\%$ error due to the fluctuations present in the simulations.
On the other hand, for $NN=10$ Steinhardt order parameter gives a few different values that are close
to one another in the range of $0.25 < q_4 < 0.325$ and $0.55 < q_6 < 0.61$.
Based on the above values, molecules with exactly ten or twelve neighbors were 
included into crystalline environments or labeled as liquid-like if any of
these conditions were not fulfilled.

\subsection{2D order parameter}
\label{sec:2d-ord}
To identify two-dimensional ordering, including discrimination between the hexagonal and
Kagom\'e structures which form alternating layers in three-dimensional tetrastack crystalline networks, 
we employed a parameter that has been proposed quite recently by H. Eslami \textit{et al.}
\cite{Kag1, Kag2}. It is defined as follows:

\begin{equation}
 \lambda_1(i)=\frac{1}{N_b(i)}\sum_{j \in N_b(i)}
 \left [ \sum_{m=-6}^6 \hat{q}_{6m} \hat{q}^*_{6m}
 - \sum_{m=-4}^4 \hat{q}_{4m} \hat{q}^*_{4m}\right ]
\end{equation}

\noindent and

\begin{equation}
 \lambda_2(i)=\frac{1}{N_b(i)}\sum_{j \in N_b(i)}
  \left [ \sum_{m=\pm6,\pm4} \hat{q}_{6m} \hat{q}^*_{6m}
 - \sum_{m=\pm4} \hat{q}_{4m} \hat{q}^*_{4m}\right ]
\end{equation}

\noindent where 

\begin{equation}
    \hat{q}_{lm}=\frac{q_{lm}(i)}
    {\left ( \displaystyle \sum_{m=-l}^l \left |q_{lm}(i) \right |^2 
    \right)^{1/2}}
\end{equation}

The hexagonal and Kagom{\'e} networks are identified in the order parameter space $(\lambda_1,\lambda_2)$
with the values of $(0.0,0.8)$ and $(0.65,0.7)$, respectively. These values correspond to the perfect lattices
and we allowed for $10\%$ uncertainty during the discrimination procedure to take into account fluctuations
present in the simulations.


\subsection{2D orientational profile}
\label{sec:2d-prof}
To get further insight into the tetrastack crystal structure, we examined how
the patchy particles are arranged in the layers with respect to the surface.
To do that, we calculate the cross product between the 
vectors pointing from the center of the patchy particle to the centers of
the two neighboring active sites. This results in a new vector $\hat{u}_i$ that is perpendicular
to the plane formed by these two vectors (green square in Figure~\ref{fig:model} a). 
Next, for each of the particles,
we calculate the dot product between the plane and
the reference z-axis (wall-direction) $\hat{e}_z$. 
The results are presented in the form of the distribution function of the angles and
the particle's z-coordinate $h(z,\alpha)$ where $\alpha=\arccos(\hat{u}\cdot\hat{e}_z)$. 
Using this definition of $\alpha$ the plane containing patches
is parallel to the substrate surface when $\alpha=0^\circ$.

\subsection{Simulation details}\label{sec:sim-det}
Molecular dynamics simulations were launched using
LAMMPS simulation package \cite{LAMMPS}. Trajectories were evolved
using the velocity Verlet algorithm, with a timestep of $\tau=0.001$. 
The temperature was controlled using the Nos{\'e}-Hoover chains
thermostat{\color{black} \cite{nhchains}} with damping factor $\tau_{NH}=10\tau$ and the number of chains equal to 3.
The systems comprised of $8100$ or $16129$ for systems under confinement 
to check for possible finite-size effects. 
The system size $L_x\times L_y\times L_z$ was equal to $34\times34\times40$ 
in $x, y, z$ directions for both system sizes, resulting in that the surface density
corresponds to the formation of up to seven and fourteen close-packed monolayers
for $8100$ and $16129$ patchy particles, respectively. 
In the slab geometry  used here, the periodic boundary conditions were applied in the 
x- and y-directions, while the z-direction was bound by the
Lennard-Jones (9,3) potential, and the reflecting wall at the bottom and the top sides, respectively.
{\color{black} Additional simulations were carried out for the simulation boxes with the distance between the walls ($L_z$) equal to $L_z=25\sigma$, $L_z=60\sigma$, and $L_z=80\sigma$,
and with the numbers of particles equal to $13456$, $24336$, and $32761$, respectively.
}
The bulk system was composed of $3375$ patchy particles at distinct densities 
ranging from $\rho=0.6$ to $\rho=1.3$. 
{\color{black} The motivation for performing simulations in the
canonical ensemble $NVT$ rather than in other ensembles
is as follows. We wanted to examine the behavior of
patchy particles when the system has a constant density. 
This of course not necessarily correspond to the coexistence
points in all cases, nonetheless it allows one for the assessment of
the stability of the formed ordered phases under such off-coexistence conditions
which also are of experimental interest.} 
Each of the systems was gradually cooled down from disordered states
{\color{black} starting at the temperatures $T$ that varied
depending on the system's density $\rho$ and surface potential $\xi$}
with an increment in temperature equal to $\Delta T=0.01$.
Once the Steinhardt order parameter indicated the nucleation event, the
increment was changed to $\Delta T=0.005$. 
Simulations were launched for $2\times10^8-1\times10^9$ simulation steps at every thermodynamic state for the equilibration period. 
Further production runs were launched for at least $10^7$ timesteps.
{\color{black} In order to ensure the results do not change from
one replica to another, we performed three independent simulation runs 
for all the cases studied. }

\section{Results and discussion}

\subsection{Phase behavior of the bulk system}

\begin{figure}[h]
    \centering
    \includegraphics[width=\linewidth]{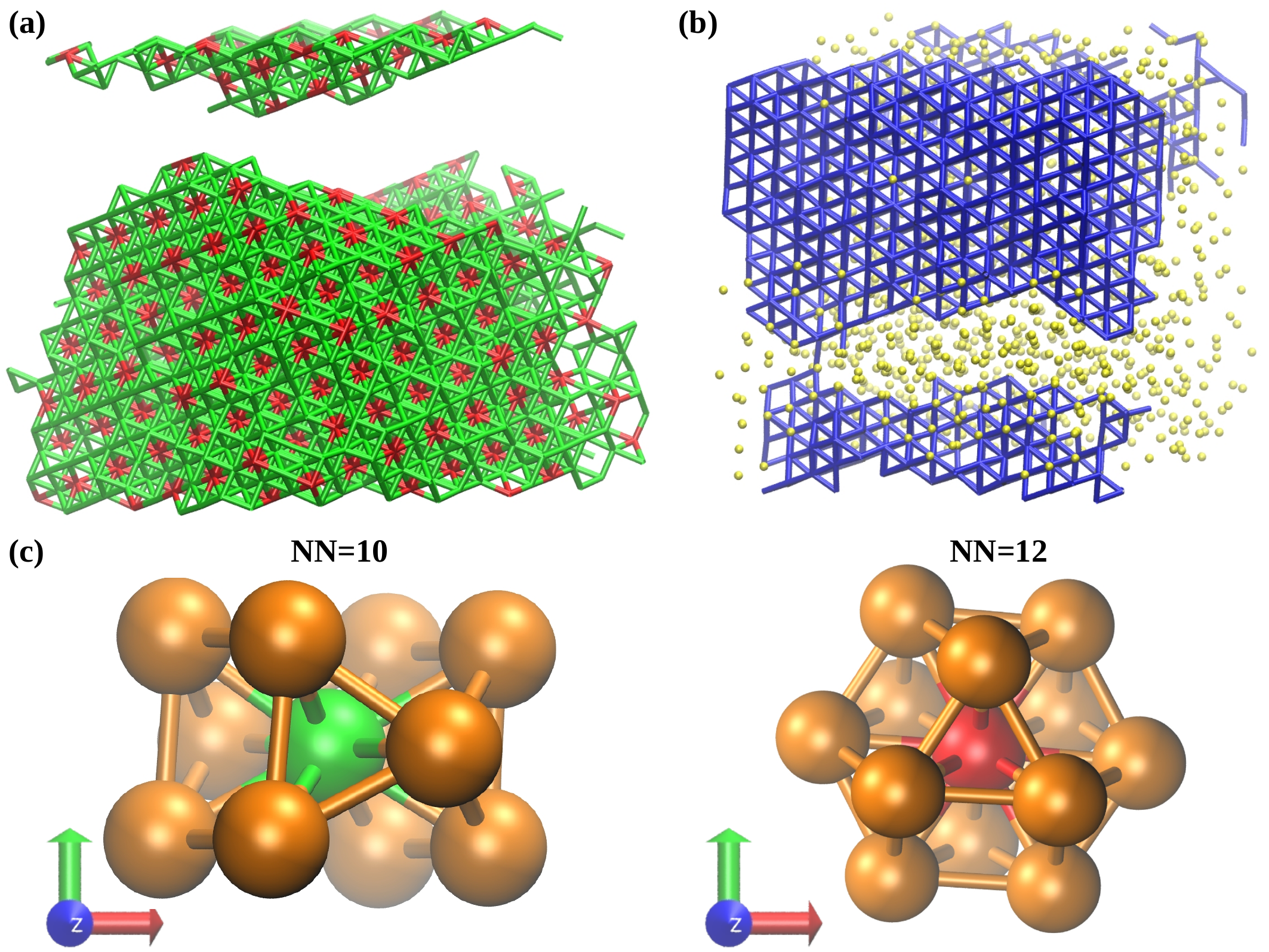}
    \caption{Snapshots demonstrating the cubic tetrastack (a)
    and fcc (b) crystalline networks. Green and red sticks correspond
    to the particles with 10 and 12 nearest neighbors, respectively.
    In part (a) fluid particles have been omitted for the sake of clarity. 
    In part (b) fcc lattice is marked as blue sticks whereas yellow spheres depict fluid particles. {\color{black} Part (c):
    Local environments with NN=10 (left) and N=12 (right)
    in cubic tetrastack crystals. Central atoms are colored as
    in part (a).}}
    \label{fig:bulk-snap}
\end{figure}

Before we turn to the main subject of the study, i.e., to the assembly of model patchy particles subjected to the
external field, it seems reasonable to begin with the discussion of the bulk behavior. To this end, we have performed extensive simulations
over a wide range of temperatures, and at different densities.
It should be emphasized that we have not attempted to evaluate the entire phase diagram, but rather to demonstrate its topology, 
and to determine the structure of different ordered phases. In particular, we have confined the  calculations to the densities up to 
$\rho =1.30$. At still higher densities, the simulations turned out to be very slowly converging, and the time needed to reach equilibrium states 
appeared prohibitively long.   

Figure~\ref{fig:bulk-snap} (a) shows the snapshot, being a representative example of configurations emerging from
the simulations at low temperatures, below $T=0.92$, and at the densities between $\rho=1.0$ and 1.2. This snapshot demonstrates 
the formation of a complex layered structure. On the other hand, at the temperatures above 0.92, the recorded configurations have shown the formation
 of large domains of crystalline phase coexisting with a disordered fluid (see Fig.~\ref{fig:bulk-snap} (b)).
 
The inspection of several configurations demonstrated that the 
low temperature ordered phase consists of alternate layers, in which the particles form the [111] face of fcc, and the Kagom{\'e} lattice, respectively. 
Although it is not seen in the snapshot given in Fig.~\ref{fig:bulk-snap} (a), 
the fcc layers are periodically shifted, since the particles in subsequent three layers (e.g., 1,3, and 5) occupy one of the three 
interpenetrating sublattices, A, B, and C. The same applies to the Kagom{\'e} layers, (e.g., 2, 4, and 6).
Thus, the ordered low temperature structure is the cubic tetrastack (CT) polymorph. 
On the other hand, the high temperature ordered phase has been found to be the fcc crystal. 

\begin{figure}
    \centering
    \includegraphics[width=\linewidth]{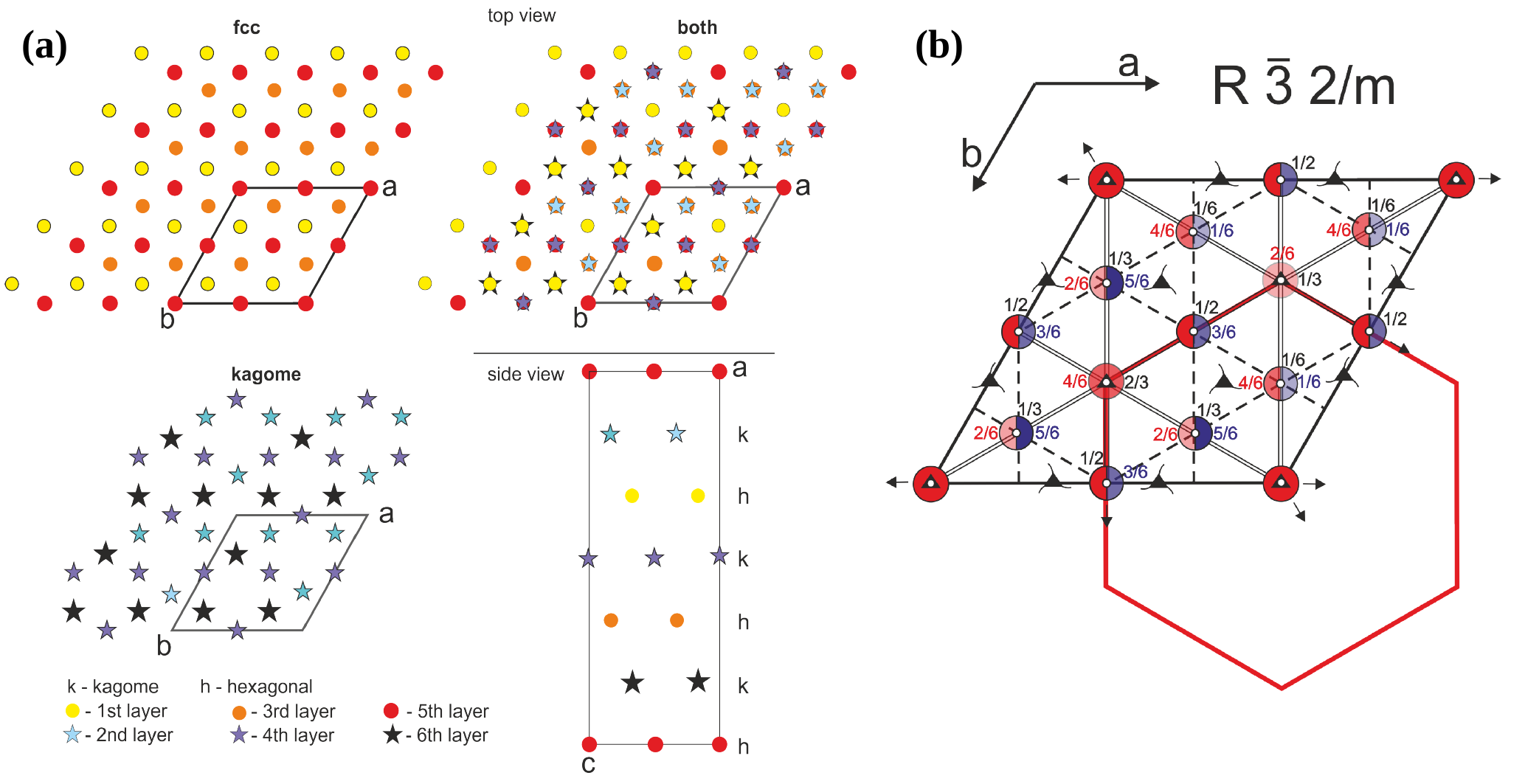}
    \caption{Part (a) demonstrates the lattice of fcc (upper left), Kagom{\'e} (lower left) 
    structures, and their composition as seen from above (upper right) and from the side (lower left). 
    Part (b): unit cell projection of the cubic tetrastack crystal belonging to the
    R$\bar{3}$2/m space group with the selected symmetry elements. }
    \label{fig:lattice}
\end{figure}

A detailed analysis of the CT phase has allowed us to construct its unit cell. 
The upper left panel of Fig.~\ref{fig:lattice} (a) presents the lattice of the fcc structure with high symmetry space group F4$\bar{3}$m, 
 seen from the [111] direction. The {\color{black} three-fold inversion axis}
can be directly noticed. Adding the Kagom{\'e} layers between the fcc layers (see the lower left panel of Fig.~\ref{fig:lattice} (a))
lowers the system's symmetry (see the right panels of Fig.~\ref{fig:lattice} (a)). One should note the periodic shifts in particle's positions in the subsequent layers 
of the fcc and Kagom{\'e} lattices. The new symmetry of the resulting CT structure is rhombohedral R$\bar{3}$ 2/m. 
In this case, the {\color{black} three-fold inversion axis} remains unchanged. Figure~\ref{fig:lattice} (b)
presents the unit cell projection with atoms' positions and their heights (colored numbers), together with 
the symmetry elements related to this symmetry group. The black numbers are the height of the symmetry elements ({\color{black} inversion centers}). 
The red hexagon in Fig.~\ref{fig:lattice} (b) represents the equivalent cell used to describe a rhombohedral system.

We have generated a perfect CT phase and found that the lattice unit cell, containing 21 particles, is characterized by the lattice vectors
$\vec{a}=(2,0,0)$, $\vec{b}=(0,2,0)$ and $\vec{c}=(0,0,4.8)$. The fractional positions of all atoms in the unit cell are summarized in Table~\ref{tab:lattice_positions}. 

\begin{table}
    \caption{The positions of atoms, in fractional coordinates, belonging to the
    unit cell of the cubic tetrastack (CT) crystal.}
    \centering
    \begin{tabular}{cccc}
        No. & $x/|\vec{a}|$ & $y/|\vec{b}|$ & $z/|\vec{c}|$ \\
        \hline
        \hline
        1 & 0.0 & 0.0 & 0.0 \\
        2 & 0.5 & 0.0 & 0.0 \\
        3 & 0.0 & 0.5 & 0.0 \\
        4 & 0.5 & 0.5 & 0.0 \\
        5 & 0.66667 & 0.83333  & 0.16667 \\
        6 & 0.16667 & 0.83333 & 0.16667 \\
        7 & 0.16667 & 0.33333 & 0.16667 \\
        8  & 0.33333 & 0.66667 & 0.33333 \\
        9  & 0.83333 & 0.66667 & 0.33333 \\
       10  & 0.33333 & 0.16667 & 0.33333 \\
       11  & 0.83333 & 0.16667 & 0.33333 \\
       12 & 0.0 & 0.5 & 0.5 \\
       13 & 0.5 & 0.5  & 0.5 \\
       14 & 0.5 & 0.0 & 0.5 \\
       15 & 0.66667 & 0.33333 & 0.66667 \\
       16 & 0.16667 & 0.33333 & 0.66667 \\
       17 & 0.66667 & 0.83333 & 0.66667 \\
       18 & 0.16667 & 0.83333 & 0.66667 \\
       19 & 0.3333 & 0.16667  & 0.83333 \\
       20 & 0.8333 & 0.16667 & 0.83333 \\
       21 & 0.3333 & 0.66667 & 0.83333 \\
        \hline
        \hline
    \end{tabular}
    \label{tab:lattice_positions}
\end{table}

In order to determine the changes in the ordering of the bulk with temperature and density, we have considered the behavior of the Steinhardt order parameters 
$q_4$ and $q_6$. For the ideal CT structure, these order parameters allow to discriminate the particles with 10 and 12 nearest 
neighbors, as discussed in Section~\ref{sec:3d-ord}.

\begin{figure}
    \centering
    \includegraphics[width=\linewidth]{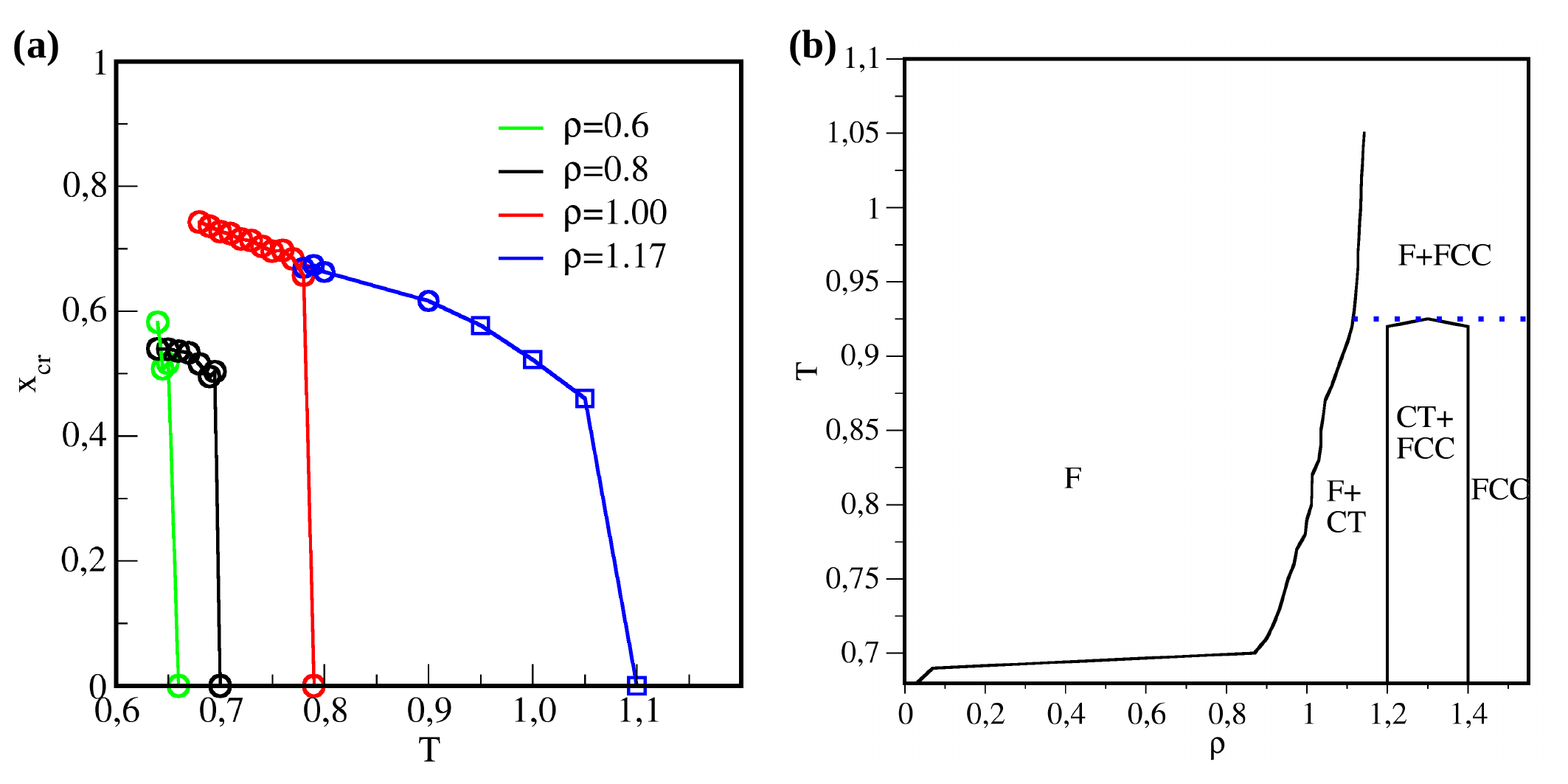}
    \caption{Part (a): Temperature relation of the ratio of crystalline particles
    to the total number of molecules $x_{cr}$. Circles and squares correspond to the cubic
    tetrastack and fcc environments, respectively. Part (b): The fragment of schematic
    bulk phase diagram in the $\rho-T$ plane. }
    \label{fig:bulk-diag}
\end{figure}

The calculations of temperature changes of $q_4$ and $q_6$
 allowed to estimate the 
contributions of particles belonging to the ordered phases, measured by the ratio of particles classified as belonging to the crystalline phase to the total number of particles,
$x_{cr} = N_{cr}/N_{total}$.
The results are given in Fig.~\ref{fig:bulk-diag} (a), and show that an increase in temperature leads to a gradual decrease of $x_{cr}$, up to the temperature, $T_{tr}(\rho)$,
at which $x_{cr}$ drops to zero. This sudden disordering suggests the presence of first-order transition. Figure~\ref{fig:bulk-diag} (a)
also shows that the transition
temperature increases with $\rho$. Besides, from the calculated order parameters, we have found that for 
$\rho\leq 1.0$, the disordering transition occurs between the CT ordered phase and the fluid, while at the density equal to 1.20, the CT   
phase is stable only at temperatures up to about 0.92. At higher temperatures, a vast majority of particles involved in the crystalline phase
have been found to have 12 nearest neighbors and form the fcc crystal. The example of configuration recorded for $\rho=1.20$, and at $T=1.05$ has already been shown 
in Figure~\ref{fig:bulk-snap} (b).

The results have allowed us to conclude that the bulk phase diagram looks like that shown in Fig.~\ref{fig:bulk-diag} (b), and its topology is qualitatively similar to 
that found by Romano {\color{black}{\it et. al} \cite{ph1} }for tetrahedral patchy particles. Thus, the system does not exhibit the gas-liquid transition, 
but only the transition between the fluid and solid phases.
Notice, however, that such an evaluated phase diagram is only schematical
as we were mainly interested in the approximate estimation of 
its topology and the coexistence regions. {\color{black} Moreover,
we cannot completely exclude the possibility of the existence of a metastable liquid phase.
To unequivocally establish its presence would require rather tedious calculations
and was beyond the scope of this article.}
At temperatures up to about 0.92, the fluid coexists with the CT ordered phase. It should be noted,
however, that the solid is not a perfectly ordered CT crystal, but contains some small fraction (between 10 to 15\%)
of the fcc structure.  At temperatures above 0.92, the solid phase exhibits the fcc structure.
It should be emphasized that at the densities equal to 1.20 and 1.30, and at the temperatures below 0.92 we have observed the presence of coexisting CT and fcc structures.
Therefore, it can be anticipated that at sufficiently high densities, the solid phase should be a pure fcc crystal. This expectation is
supported by the high temperature behavior, which demonstrates that only the fcc structure coexists with the disordered liquid. Therefore, one expects that upon cooling
the system at sufficiently high density should form a pure fcc crystal. It has to be emphasized, however, that the location of the line delimiting the region of 
densities over which the CT and fcc structures coexist, from the high density region in which only the fcc structure occurs, has not been estimated. Nonetheless, 
it is bound to appear at the density higher than 1.3.

\subsection{Ordering in the external field}
 
 \begin{figure}[h!]
    \centering
    \includegraphics[width=\linewidth]{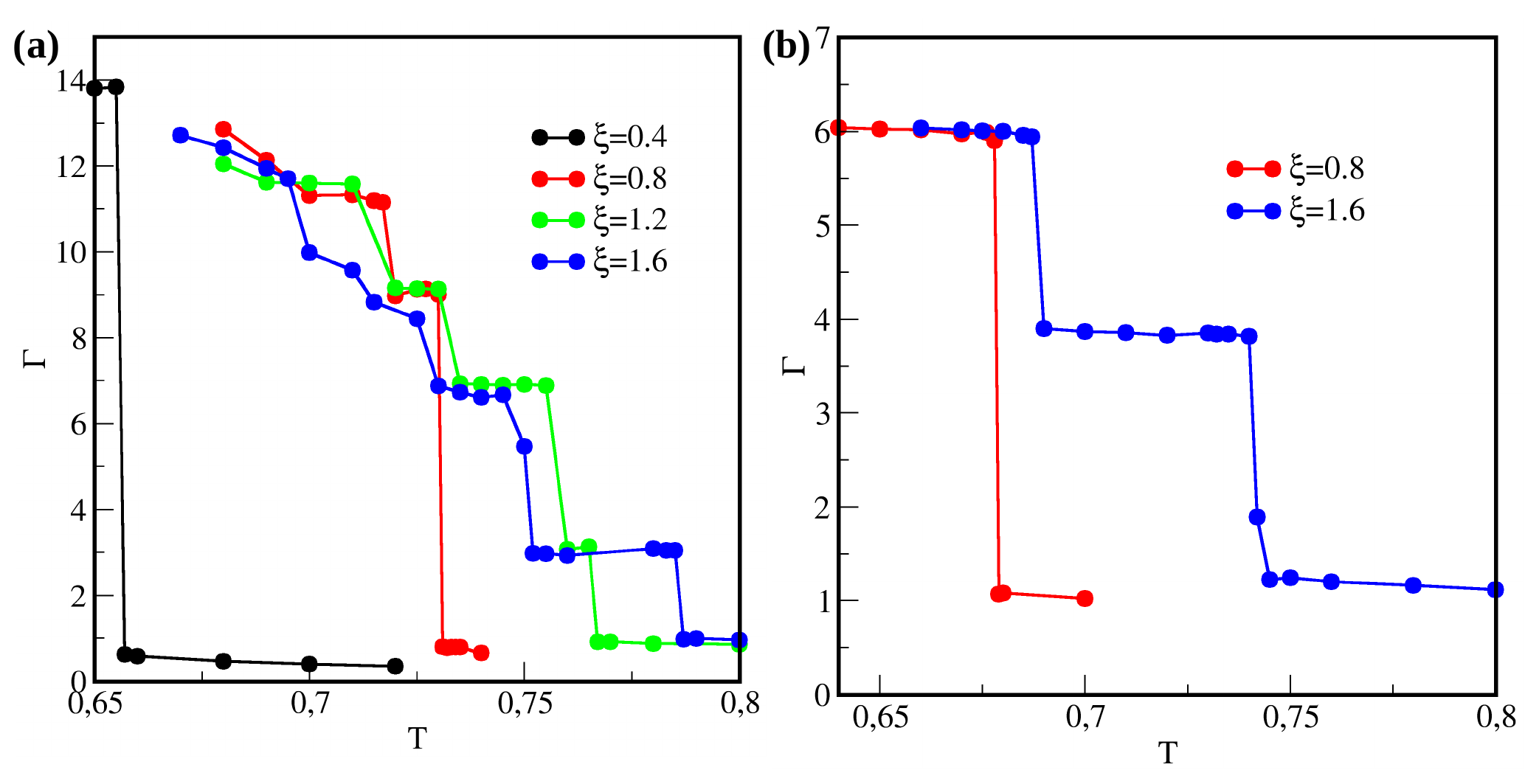}
    \caption{Changes of the surface excess density $\Gamma$ with temperature
    for systems with the total density equal to $\rho=0.352$ (a) and $\rho=0.173$ (b). }
    \label{fig:external-gamma}
\end{figure}
 
It is well known that the formation of adsorbed films is primarily determined by the relative strength of the particle-particle ($\epsilon_{aa}$) 
and the particle-wall ($\epsilon_{wc}$) interactions\cite{AD1}. Here,  we have studied
self-assembly of patchy particles subjected to the external field of varying strength, measured by the parameter $\xi$ (cf. Section~\ref{sec:patch}). 
We have considered four different values of $\xi= 0.4$, 0.8. 1.2 and 1.6, and performed simulations at different temperatures, and for
two different values of the total density equal to $\rho=0.173$ and 0.352. 

In order to study the development of adsorbed films, the  
density profiles, $\rho(z)$, have been recorded, and used to  
calculate the surface excess densities, $\Gamma$,
\begin{equation}
 \Gamma = \frac{1}{S}\int_0^{L_{z,max}}[\rho(z)-\rho_b]dz
 \label{eq:gamma}
\end{equation}
In the above, $S=L_xL_y$ is the surface area, $L_{z,max}=35$, and $\rho_b$ is the bulk density. The bulk density has been calculated as
\begin{equation}
 \rho_b = \frac{1}{S(L_{z,max}-L_{z,min})}\int_{L_{z,min}}^{L_{z,max}}\rho(z)dz
 \label{bulk}
\end{equation}
with $L_{z,min}= 25$. The selected value of $L_{z,min}$ is beyond the range of $z$, over which the density profiles have shown the peaks associated with the formation of adsorbed layers. At the same time $L_{z,max}=35$ ensures that
any undesired effects stemming from the presence of the reflecting wall are avoided. 
  
The central results of our calculations are given in Figure ~\ref{fig:external-gamma}, which presents the changes in the surface excess density with temperature, for different values of
$\xi$ (cf. Section~\ref{sec:patch}), and for $\rho=0.352$ (part (a) of Fig.~\ref{fig:external-gamma}) and $0.173$ (part (b) of Fig.~\ref{fig:external-gamma}). In all cases, the surface excess density exhibits a series of jumps, which 
indicate the formation of a gradually increasing number of adsorbed layers in the film, when the temperature drops. 
The sequence of these layering transitions changes when the strength of surface potential increases, and depends also on the total density.

\begin{figure}
    \centering
    \includegraphics[width=\linewidth]{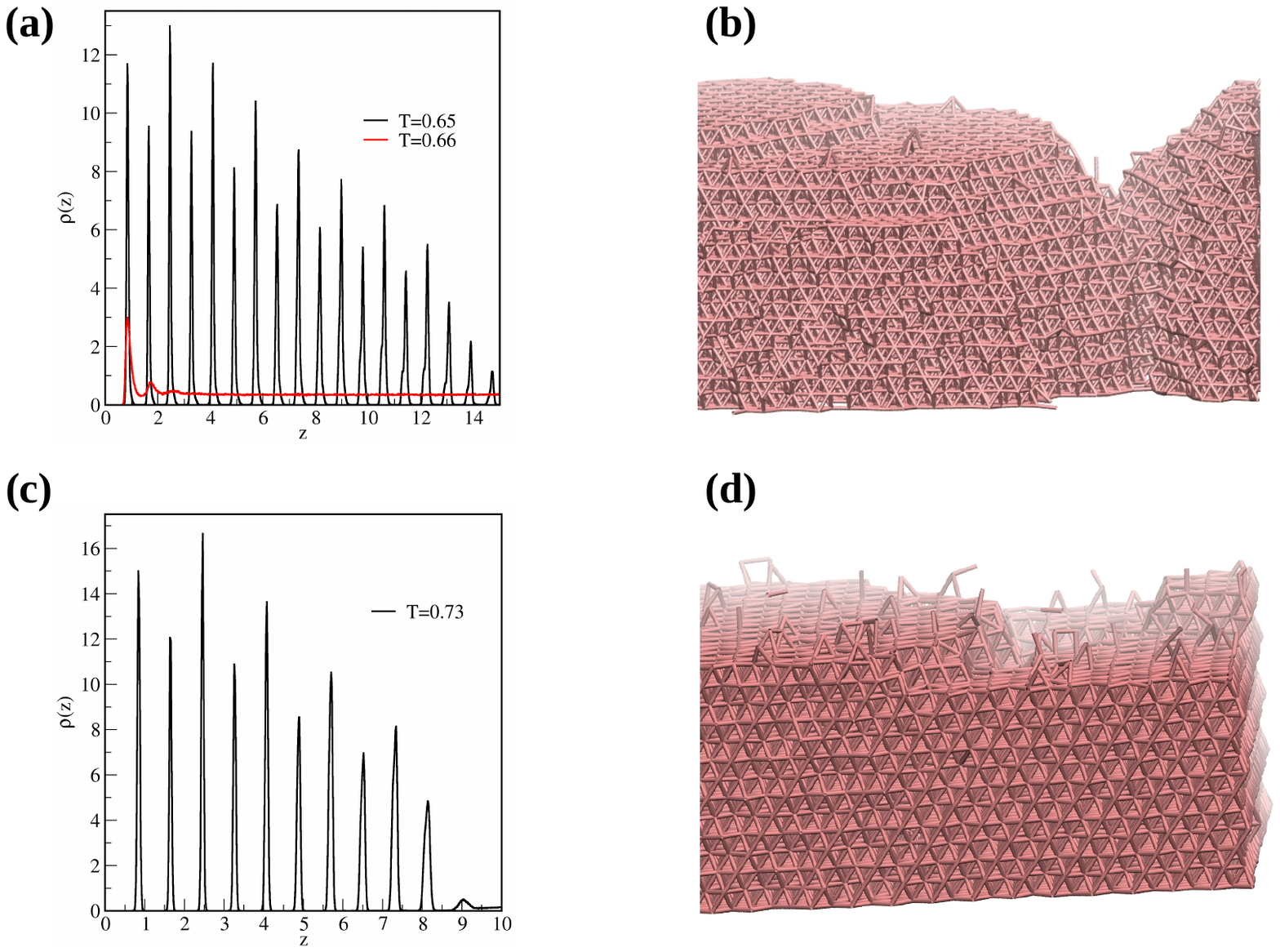}
    \caption{Density profiles for the systems with $\xi=0.4$ (a) and $\xi=0.8$ (c). 
    The corresponding snapshots recorded at $T=0.65$ (b) and {\color{black}$T=0.68$} (d)
    for parts (a) and (c), respectively. The system density was equal to $\rho=0.352$ in all cases. 
    }
    \label{fig:profiles1}
\end{figure}

In the case of the weakest surface potential, with $\xi=0.4$, we have performed the simulation for the bulk density $\rho =0.352$, and found only one 
transition leading to the formation of a very thick film. 
Figure~\ref{fig:profiles1} (a) shows the density profiles recorded at the temperatures  below ($T=0.65$) and above ($T=0.66$) the transition leading to the condensation of 
particles, and the formation of the thick film. 
 It is evident that at $T=0.66$ only a fraction
 of the first layer is covered, while a vast majority of particles remains in the uniform bulk phase. On the other hand, at $T=0.65$, the film consists of 18 layers,
 while the bulk density is very low. Thus, the transition may be treated as the bulk condensation, and such behavior is characteristic of the systems that do not exhibit
 wetting. 
 It should be noted that the heights of subsequent peaks corresponding to odd layers, 1, 3, $\ldots$, are considerably higher than those corresponding to even 
 layers, 2, 4, $\ldots$. and that the heights of all maxima gradually decrease with $z$. Nonetheless, the integration of density profiles has shown that the ratio
 of densities in neighboring odd and even layers is equal to $1.34\pm 0.03$ across the entire film. This suggests that the film ordered into the  
 CT structure, in which the ratio of densities of the fcc and Kagom{\'e} layers  is equal to $4/3\approx 1.33$.   
 A gradual decay of subsequent maxima at the density profile demonstrates that the 
 interface between the ordered film and the bulk dilute (gas)
 phase is rough, since the filling of higher layers is only partial (see the snapshot given in part b of Fig.~\ref{fig:profiles1}). 
 
 In the case of the stronger surface potential, with $\xi =0.8$, and $\rho =0.352$, the first transition, at $T\approx 0.73$, leads to the formation of
 10 filled layers (see the density profile given in Fig.~\ref{fig:profiles1} (c)), and is followed by the transition, at $T\approx 0.72$, leading to the mutual condensation of the next two layers.
 A further lowering of temperature, down to $T=0.68$, causes a gradual increase in the film thickness, but the interface between the film and the bulk
 is not so rough as in the case of $\xi=0.4$.  This is illustrated by the snapshot given in Fig.~\ref{fig:profiles1} (d), and recorded for the system with $\xi=0.8$, and at $T=0.68$.
 The results given in Fig.~\ref{fig:external-gamma} (a) clearly demonstrate that the mode of film development is strongly affected by the strength of the surface potential. 
For the still stronger surface fields, with $\xi=1.2$ and 1.6, the first layering transition involves a mutual condensation of only four layers,
on top of already existing monolayer film. 
 
 In the system with the lower total density, equal to $\rho =0.173$, and $\xi =0.8$ there is only one transition, 
 which leads to the formation of 6 filled layers (see part b of Fig.~\ref{fig:external-gamma}).
The film thickness is lower than in the case of $\rho=0.352$, since the number of particles in the system allows for the formation of only 6 layers. 
However, the density profile
recorded for the system with $\rho =0.173$ at $T=0.65$ demonstrates quite similar behavior to that found for thick films in the system with $\rho =0.352$.
In particular, the ratio of densities in the odd and even layers is practically the same as in the system with $\rho=0.352$,  
and implies that the film structure is also the same.  
 
 It should be noted that in all systems with $\xi\geq 0.8$, quite dense, and disordered, monolayer films are formed at the temperatures
above the first layering transition. However, the first layering transition induces ordering in the entire film, and, again, the odd layers assume the structure corresponding to
the [111] face of fcc crystal, while the even layers are ordered into the Kagom{\'e} lattice. A gradual decrease of temperature leads to a series of layering transitions, 
involving a mutual condensation of an even number of layers. This behavior can be attributed to the growth of CT phase, which retains the highest 
stability when the numbers of layers with hexagonal and Kagom{\'e} orderings are the same. We recall that the first layer exhibits hexagonal ordering, hence the top layer is bound
to order into the Kagom{\'e} lattice.

In order to confirm that thick films are ordered into the CT phase we have calculated the two-dimensional order parameters $\lambda_1(i)$, and $\lambda_2(i)$ 
for subsequent layers, $i$ of thick films (cf. Section~\ref{sec:2d-ord}). The regions of $z$ corresponding to each layer have been estimated from the density profiles, 
and these order parameters allowed us to estimate the contributions of
particles involved in the Kagom{\'e} and hexagonal lattices. A representative example of the results is given in Table~\ref{tab:percent}, which demonstrates that odd layers 
are well ordered into the hexagonal lattice, while even layers are ordered into the Kagom{\'e} lattice.

\begin{table}
 \caption{The percentage of particles ordered on hexagonal and Kagom{\'e} lattices in the system with the bulk density $\rho=0.352$, $\xi=0.4$, and at $T=0.655$.}
 \vspace{0.3cm}
 \centering
 \begin{tabular}{|c|cc|c|c|}
 \hline
 layer ($i$) & $z_{min}(i)$ & $z_{max}(i)$ &  \% Kagom{\'e}  & \% hexagonal \\
 \hline
 1 & 0.6 & 1.2 & 0.0  & 93.58 \\
 2 & 1.3 & 1.7 & 75.19 & 0.0 \\
 3 & 2.2 & 2.7 & 0.0 & 82.86 \\
 4 & 3.0 & 3.4 & 91.46 & 0.0 \\
 5 & 3.8 & 4.4 & 0.70 & 78.68 \\
 6 & 4.7 & 5.2 & 91.47 & 0.0  \\
 7 & 5.4 & 6.0 & 0.46 & 85.24 \\
 8 & 6.2 & 6.8 & 93.62 & 0.0  \\
 9 & 7.1 & 7.5 & 0.16 & 84.16 \\
 10 & 7.9 & 8.3 & 92.57 & 0.0 \\
 11 & 8.7 & 9.1 & 0.09 & 86.84 \\
 12 & 9.5 & 10.0 & 93.17 & 0.0  \\
 \hline
  
 \end{tabular}
\label{tab:percent}
\end{table}

\begin{figure}
    \centering
    \includegraphics[width=\linewidth]{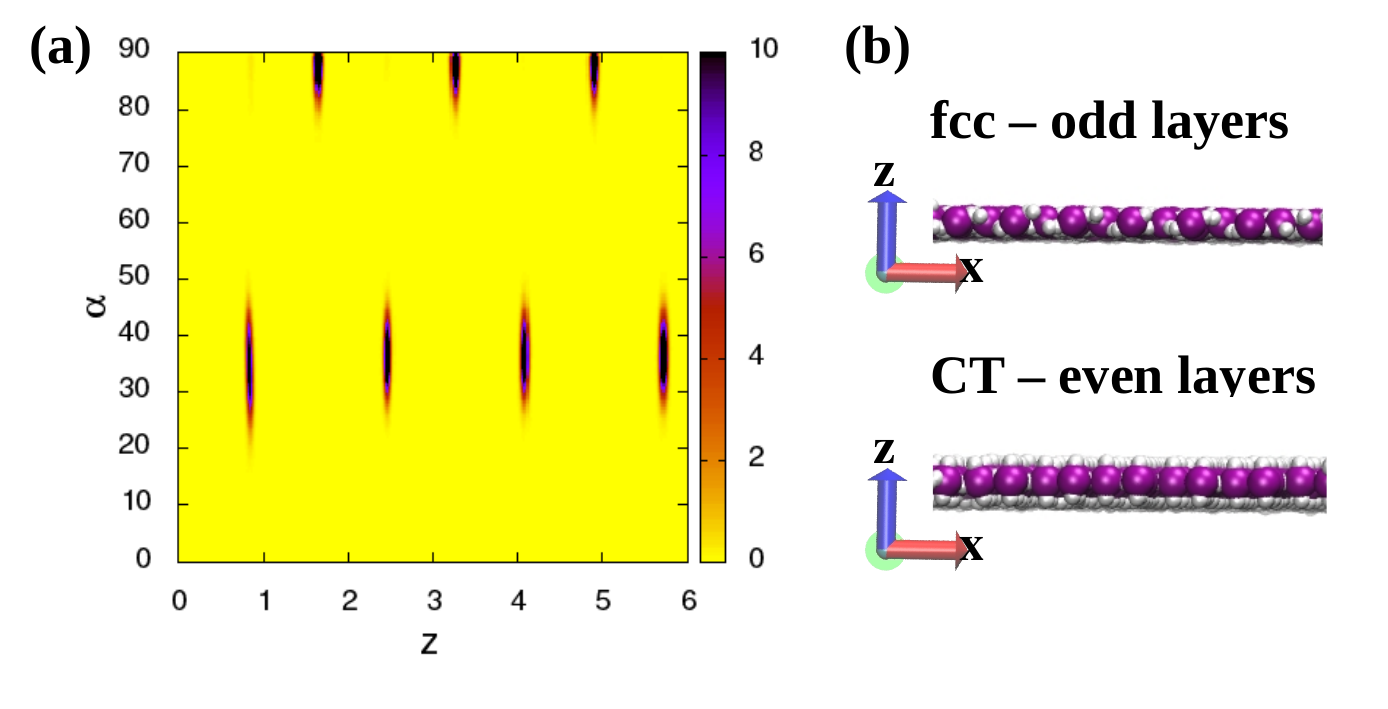}
    \caption{Two-dimensional orientational profile, $h(z,\alpha)$,
    recorded for system's density $\rho=0.352$ 
    at $T=0.68$ and $\xi=1.6$. 
    {\color{black} Part b: Fragments of fcc (odd layers)
    and CT (even layers) formed by patchy particles. The cores of
    patches and active sites are colored in purple and white, respectively.
    }}
    \label{fig:2dorder}
\end{figure}

Our model assumes that all four attractive patches are arranged in a plane. The calculations of two-dimensional orientation profiles 
(cf. Section~\ref{sec:2d-prof})  have revealed that in even layers, i.e., those forming the Kagom{\'e} lattice, the planes containing the patches are predominantly oriented perpendicular to
the substrate surface, while in the case of hexagonally ordered odd layers, those planes are tilted with respect to the surface by the angle $\alpha\ = 35\pm 10^\circ$. 
This is quite well illustrated by the behavior of the distribution function $h(z,\alpha)$, recorded for the system with $\rho =0.352$ and $\xi= 1.6$, at $T=0.68$,
and shown in Figure~\ref{fig:2dorder}.

Although we have not attempted to calculate the distribution function $h(z,\alpha)$ in the bulk, nevertheless it is very likely that it behaves in the same way.
It should be remembered that the evaluation of $h(z,\alpha)$ is done with respect to the plane dependent on the orientation of the crystal. In the bulk phase, 
the orientation of the crystal evolves during the run, hence the calculation of the distribution function $h(z,\alpha)$ would require a continuous adjustment 
of the crystal orientation relative to the chosen reference plane. Such calculations are possible, of course, but would need very time consuming calculations.

\begin{figure}
    \centering
    \includegraphics[width=0.8\linewidth]{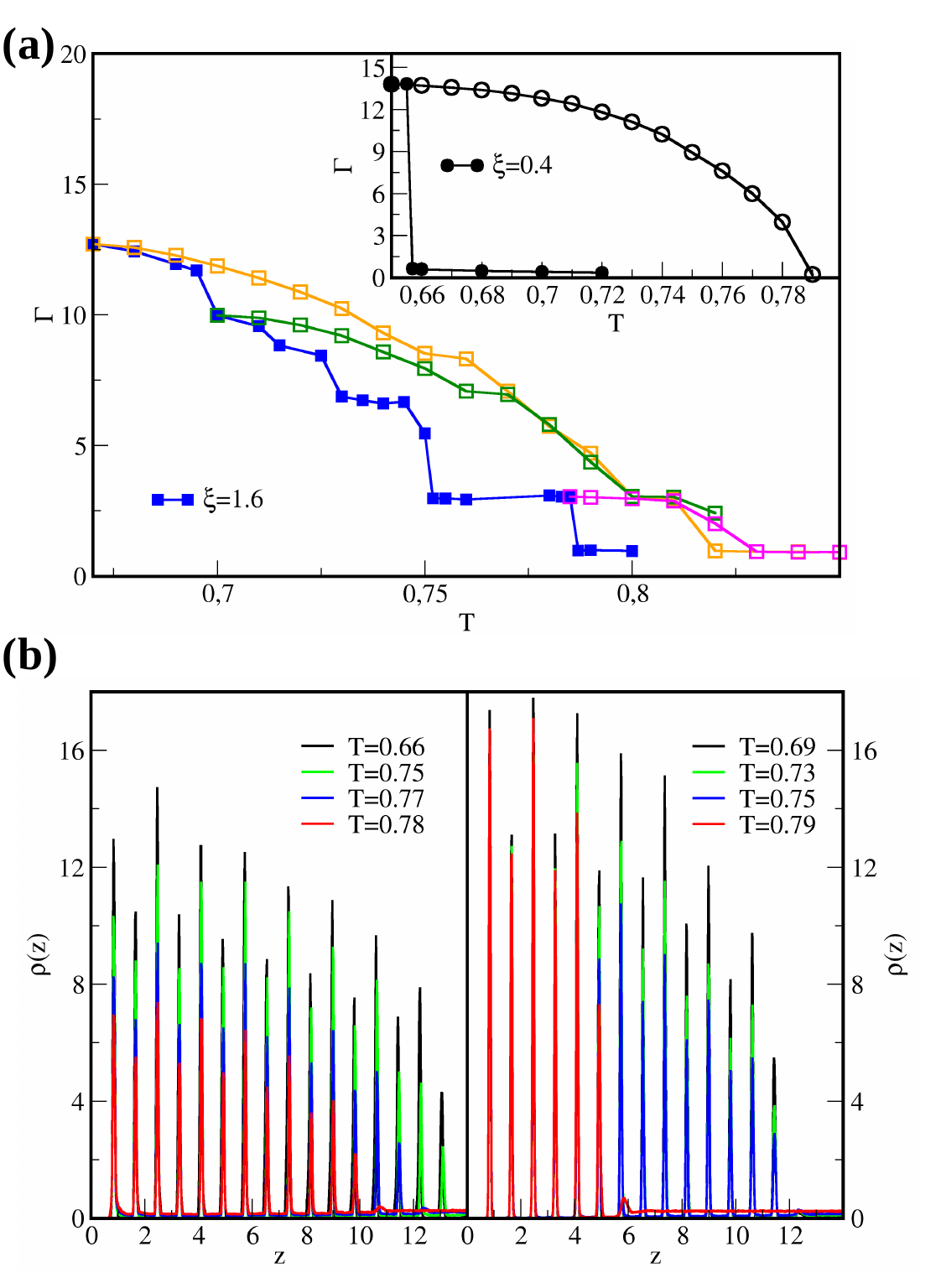}
    \caption{Part (a): Changes of the
    surface excess density $\Gamma$ with temperature for systems with the total density $\rho=0.352$ and $\xi=1.6$
    and $\xi=0.4$ (inset to this Figure).
    {\color{black} Filled and open symbols correspond to the
    condensation and evaporation simulations, respectively. In the main part of
    panel (a), starting temperatures for evaporation simulations were
    $T=0.67$ (orange curve), $T=0.7$ (green curve), and $T=0.785$.}
    Part (b): Density profiles along the evaporation curves at different temperatures
    for the systems with $\xi=0.4$ (left-hand side panel) and $\xi=1.6$ (right-hand side panel).}
    \label{fig:external-duzy}
\end{figure}

In order to study the stability of the developed films ordered into the CT phase, we have performed simulations by gradually increasing the temperature, and using the configurations 
corresponding to the  films of different thicknesses. In Figure~\ref{fig:external-duzy} (a), we present the results for the systems with $\xi=0.4$ and 1.6, and the bulk density equal to $\rho = 0.352$.
In the case of $\xi=0.4$, the only possible starting point is the film with the surface excess density
equal to $\Gamma\approx 13$, appearing at $T= 0.655$ (see the inset to Fig.~\ref{fig:external-duzy} (a)), 
while in the system with $\xi=1.6$, we have used configurations with different numbers of occupied layers (see the main part of Fig.~\ref{fig:external-duzy} (a)). 

\begin{figure}
    \centering
    \includegraphics[width=\linewidth]{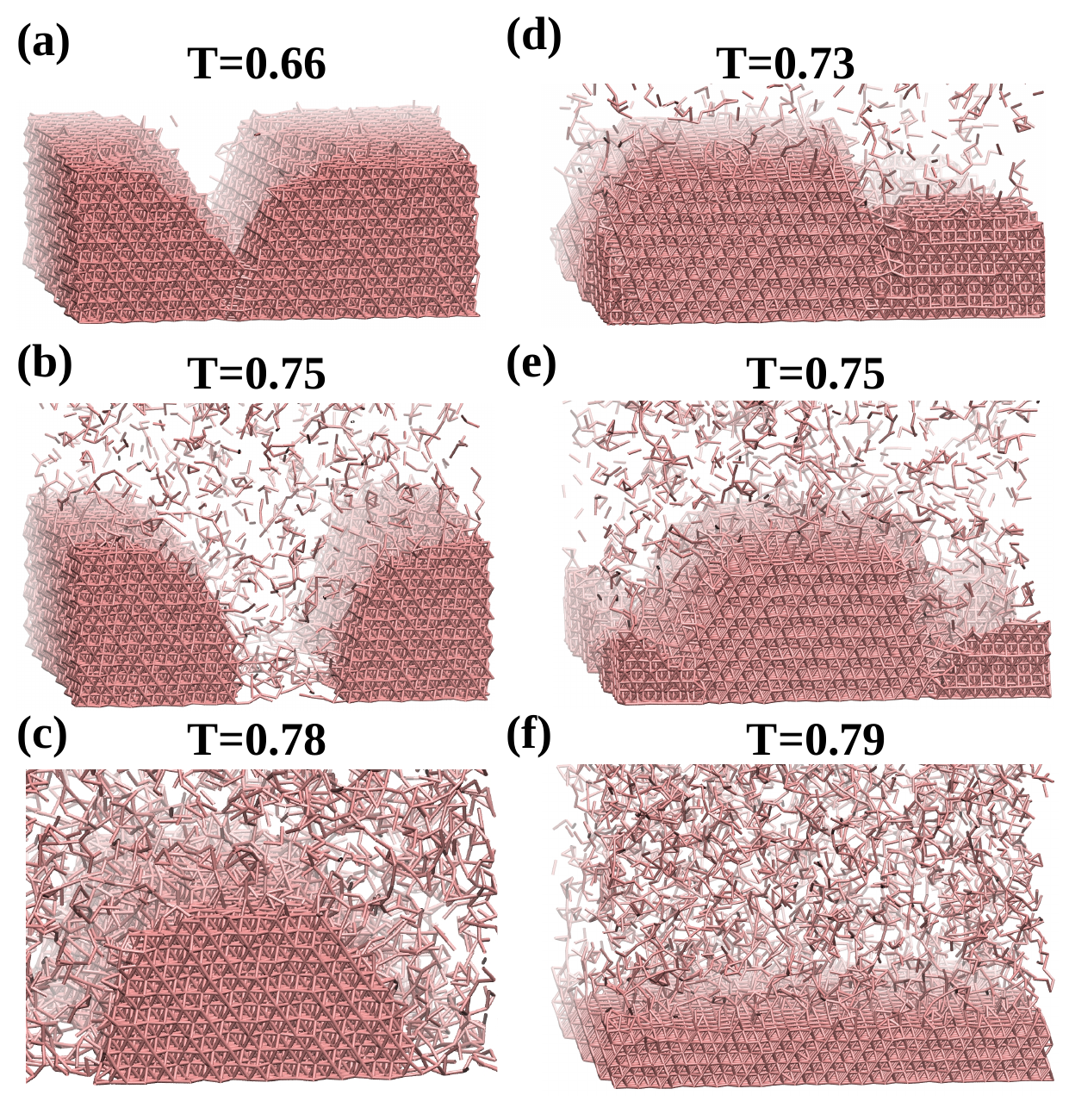}
    \caption{Snapshots along the evaporation curves at different temperatures
    for the systems with $\xi=0.4$ (a-c) and $\xi=1.6$ (d-f)
   {\color{black} at the total density $\rho=0.352$.}}
    \label{fig:snap-evaporation}
\end{figure}

The results given in Fig.~\ref{fig:external-duzy} (a) demonstrate that in both cases of $\xi =0.4$ and 1.6, the 
runs starting at high surface excess densities lead to a smooth decay $\Gamma$ when the temperature increases, and result in wide hysteresis loops. However, the mechanism 
of the film disordering depends on the external field strength. In the case of weak filed, with $\xi=0.4$, the density profiles, shown in the 
left-hand panel of Fig.~\ref{fig:external-duzy} b, demonstrate that upon 
the increase of temperature the densities of all layers gradually decrease, but even at the temperature of 0.77, there are 16 partially occupied layers. 
Only when the temperature becomes closer to the total disordering
of the film, which occurs at $T\approx 0.8$, the film thickness starts to decay.

On the other hand, the density profiles obtained for the system with $\xi=1.6$, given
in the right-hand panel of Fig.~\ref{fig:external-duzy} b, indicate that the densities of the layers close to the surface are only slightly affected by the 
increase of temperatures, and the disordering involves the top layers mostly, and leads to a gradual decrease of the film thickness. At the temperature of $T=0.75$ 
the film still consists of 14 layers, while at $T=0.79$ it has only 6 layers. A further increase of temperature to 0.8 leads to the disordering of two top layers, and 
the four-layer film remains stable up to the temperature of about 0.82. 

Figure~\ref{fig:snap-evaporation} shows the examples of snapshots recorded at gradually increasing temperatures for the systems with $\xi=0.4$ (parts a-c) and $\xi=1.6$ (parts d-f), {\color{black} starting from $T=0.65$ and $T=0.66$, respectively.}
In both systems, the disordering spreads at the interface of the crystal and fluid, but in different ways. In the case of a weak surface field, the roughness of that interface
causes that the region of disordering is primarily determined by the roughness of the crystal surface and may go down to the substrate surface, while the partially filled 
and ordered layers retain high thickness. This causes a gradual decrease of all peaks of density profiles when temperature grows. On the other hand, a strong surface field pins 
the film to the substrate and causes that the first six layers are completely filled even at quite high temperatures.
Thus, the disordering of the upper part of the film occurs in the same way as in the case of a weak surface field, and only the layers adjacent to the substrate surface exhibit higher stability.

Wide hysteresis loops between the runs corresponding to the decreasing and increasing temperature
suggest that once formed multilayer ordered structure remains stable at quite high temperatures. Does it represent truly stable states or is it just a 
manifestation of metastability? A possible way to clarify this question would be the calculation of the free energy changes along the paths of 
decreasing and increasing temperature using the method of thermodynamic integration\cite{ph2,free2}. 
Such calculations are feasible but would require the knowledge about the free energies at the certain ``reference'' points, and much computational effort. 

\subsection{Removal of the external field potential}

\begin{figure}
    \centering
    \includegraphics[width=0.75\linewidth]{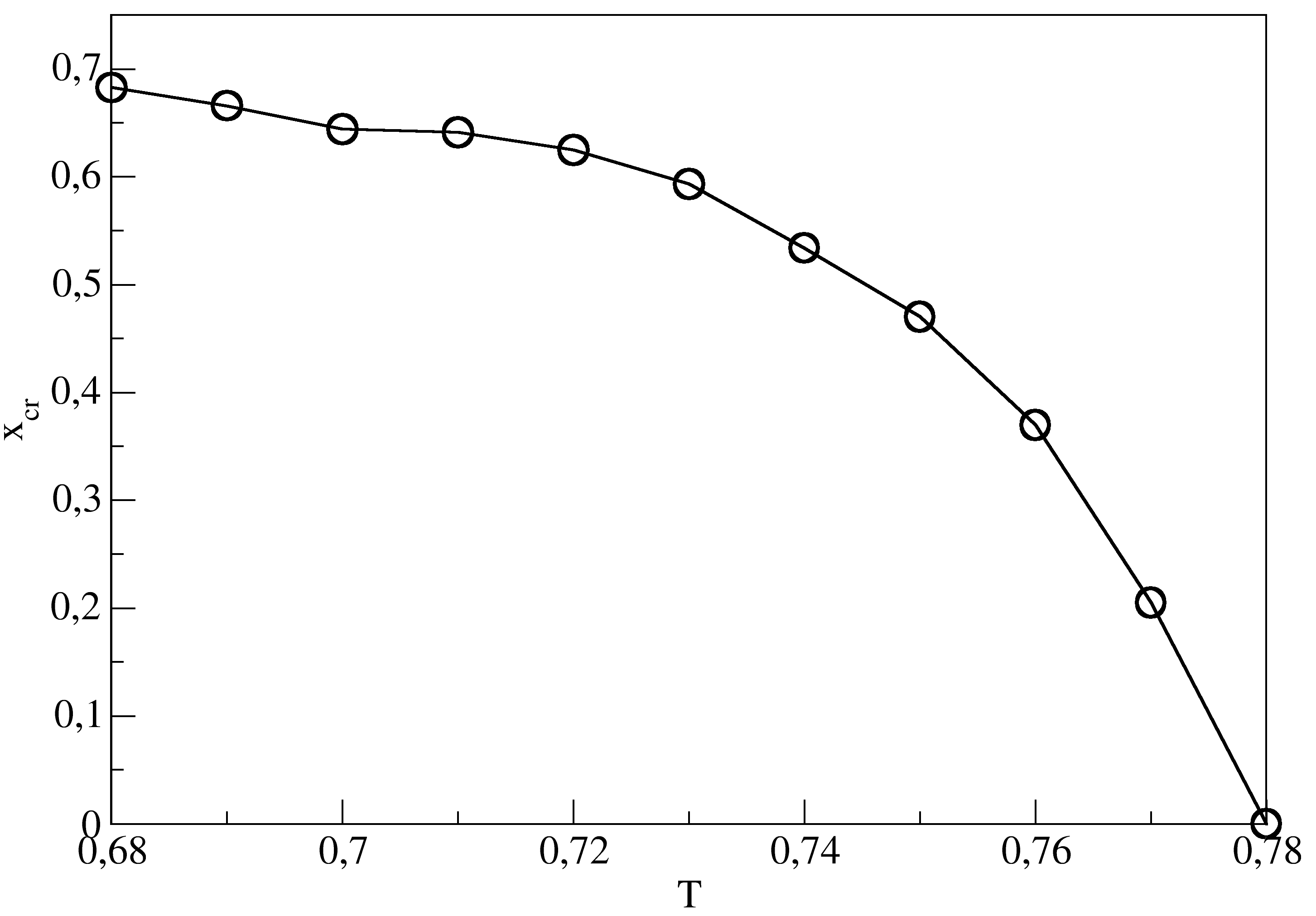}
    \caption{Temperature changes of the ratio of crystalline particles
    to the total number of molecules $x_{cr}$ when the external field
    is turned off. The starting point was $T=0.67$ and $\xi=1.6$ at the 
    systems' density $\rho=0.352$.}
    \label{fig:external-removed}
\end{figure}

\begin{figure}
    \centering
    \includegraphics[width=\linewidth]{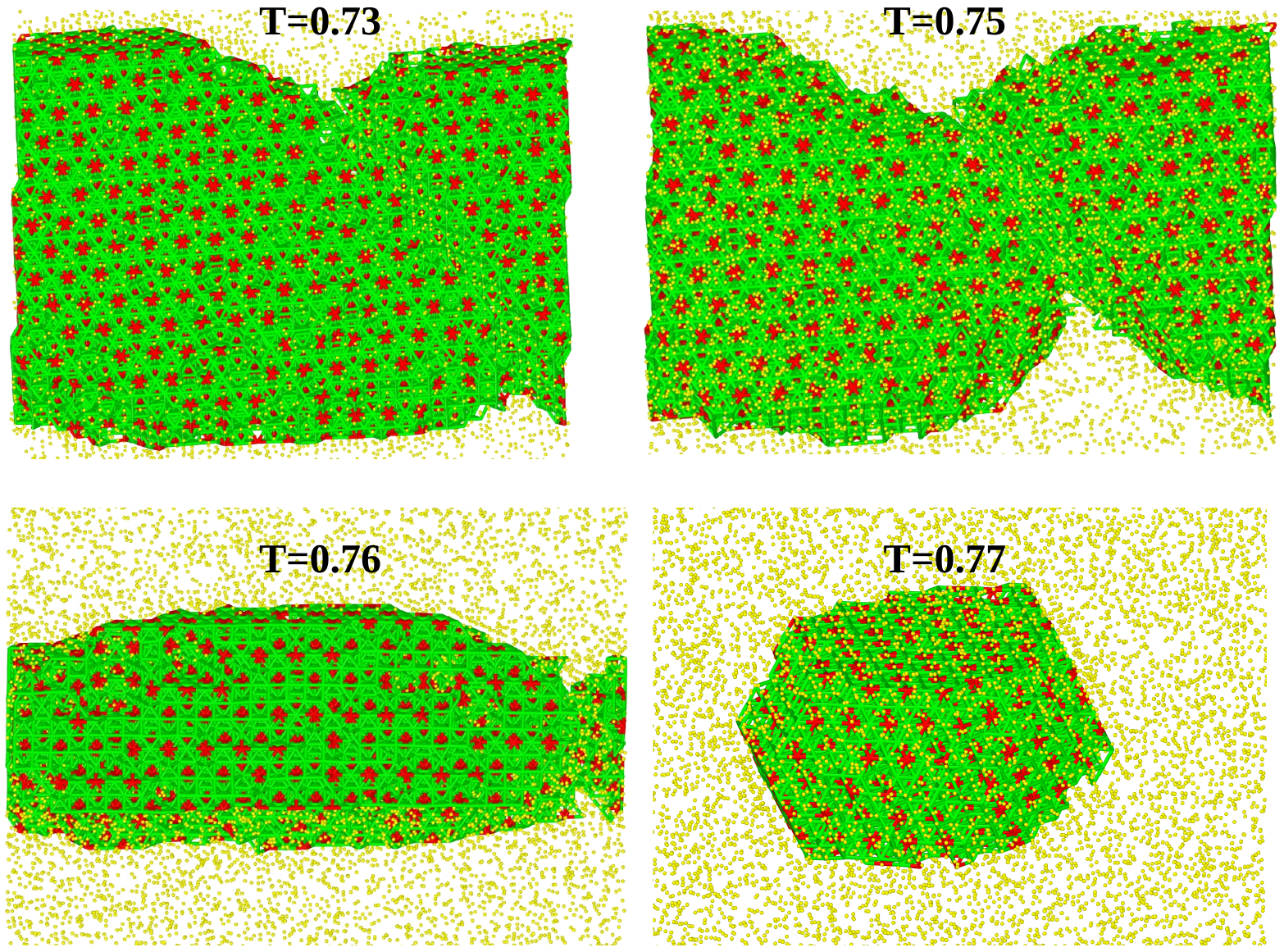}
    \caption{The examples of snapshots at different temperatures 
    for the systems with external field turned off. The starting point was $T=0.67$ and $\xi=1.6$ at the 
    systems' density $\rho=0.352$.
    Green and red sticks correspond
    to the particles with 10 and 12 nearest neighbors, respectively.}
    \label{fig:snap-external}
\end{figure}

A rather convincing evidence of the intrinsic high stability of CT crystal is provided by the results obtained when the external field was switched-off.
We have used the configuration obtained for the system with $\xi=1.6$ at $T=0.67$, with the surface excess density $\Gamma\approx 13$, and calculated 
the changes of the fraction of particles involved into the ordered CT phase when the temperature increases. The results are given in Fig.~\ref{fig:external-removed}, and 
demonstrate that  
the system undergoes a gradual disordering when the temperature increases up to $T\approx 0.78$. {\color{black} It has to emphasized that in the initial state, taken from simulations in 
the presence of external field, the system is not  in
the bulk solid-vapor coexistence. Therefore, 
a partial melting of the solid is bound to occur until the coexistence pressure
of a vapor phase is reached. The same scenario will emerge upon the increase of temperature,
until the entire crystal eventually melt in $T\approx0.78$.  
Figure~\ref{fig:snap-external} presents the snapshots recorded at 
gradually increasing temperatures. }

This finding provides a rather strong evidence that the ordered CT phase retains stability at a wide range of temperatures, and {\color{black} vanishes} only when the temperature approaches about 0.78. 
Recall that the bulk system of the density $\rho\approx0.352$ disorder at considerably lower 
temperatures, below $T\approx0.7$ (cf. Fig.~\ref{fig:bulk-diag} (b)). 
{\color{black} However, as already stated, in order to unambiguously 
determine the stability of the CT phase, rigorous free energy calculations are
required. Nevertheless, even if it were not thermodynamically stable
the appearance of such a crystal can be observed in both simulations and 
experiments.}

{\color{black}
\subsection{The effects due to the simulation box size in the z-direction}

\begin{figure}
    \centering
    \includegraphics[width=0.7\linewidth]{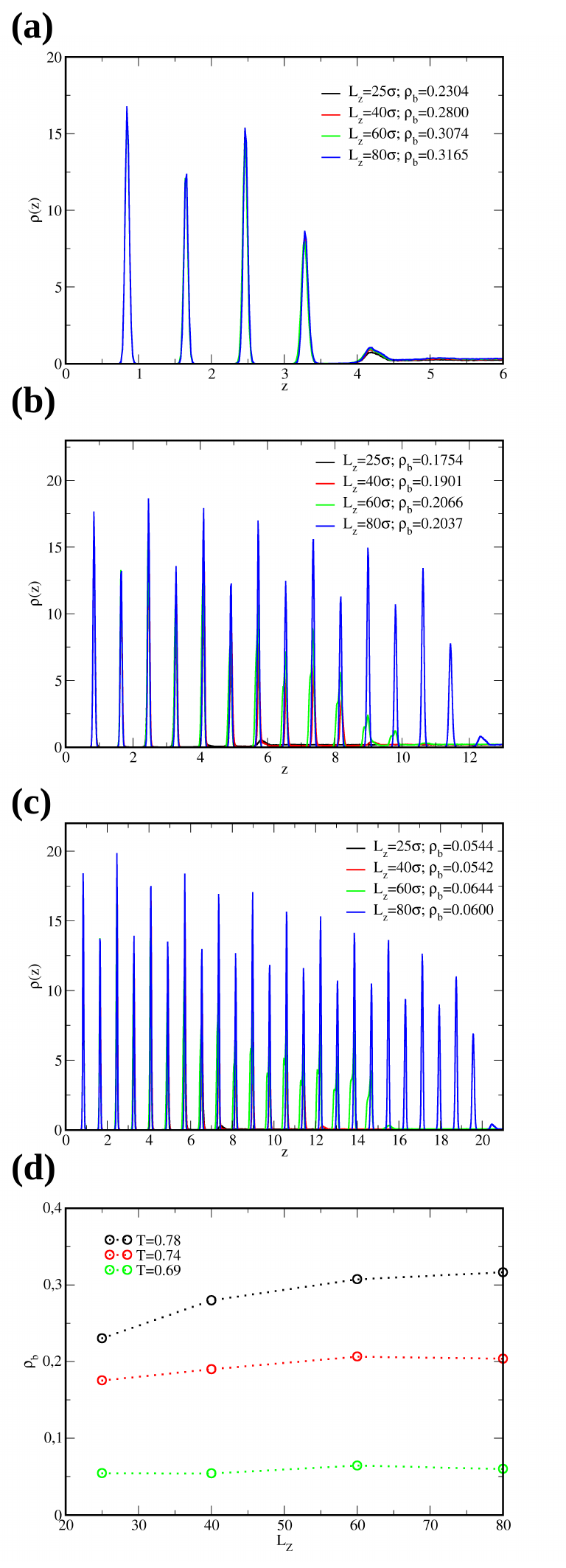}
    \caption{{\color{black}
    The comparison of density profiles evaluated at $T=0.78$ (a),
    $T=0.74$ (b), and $T=0.69$ (c) for the separation distance between 
    the walls $L_z=25\sigma,40\sigma,60\sigma$ and $L_z=80\sigma$ for a system 
    with $\xi=1.6$. The legend also 
    displays values of the bulk density $\rho_b$.
    Part (d): The relation of bulk density $\rho_b$ with the separation distance $L_z$.}
    }
    \label{fig:lz}
\end{figure}

 {\color{black}
Since we have performed the simulations in the $NVT$ ensemble
it is reasonable to verify the importance of the plausible effects
stemming from the insufficiently large system size in the $z$-direction.
A comparison of the results obtained for the system 
with the total density equal to $\rho=0.352$ and different $L_z$ is shown in Figure~\ref{fig:lz}.
As the separation distance between the walls becomes larger, 
the developing crystalline layer can be thicker. This effect becomes more pronounced
when the temperature becomes lower (cf. Figure~\ref{fig:lz} c). 
Nevertheless, it is evident that the growth of crystalline network
still occurs in a layer-by-layer fashion. The number of growing layers
is always even, which is consistent with the proposed mechanism 
of the crystallization of the cubic tetrastack phase.

One should also notice a slight increase of the bulk density, $\rho_b$, when
$L_z$ becomes larger, and it occurs at all temperatures considered, as shown in Figure~\ref{fig:lz} (d). 
However, the differences in $\rho_b$  gradually decrease
as the size of the simulation box becomes larger, and are expected to vanish
for sufficiently large $L_z$. In fact, the results at $T=0.69$ and 0.74 demonstrate that 
for $L_z=60$ and 80 the bulk density is the same, within statistical error limits.
We conjecture that the changes in $\rho_b$ with $L_z$ can be attributed 
to the finite size of the system and finite length of simulations in the $NVT$
ensemble. Together with the increase in $L_z$, the number of particles in the system of the same total density
is also larger, resulting in the necessity of sampling much larger phase space in order
to obtain reliable results, and so the simulation time becomes prohibitively long.

The above described behavior is an artifact of simulations in the $NVT$ ensemble.
Having the total number of molecules fixed, we have a limited reservoir of 
particles able to adsorb on the surface. Upon the increase of $L_z$,
while keeping the same total density, the number of molecules increase which allows for the formation of higher
film, as shown in Figure~\ref{fig:lz}. This could be potentially alleviated by
perfoming the simulations in the grand canonical ensemble $\mu VT$. However,
we have already explained the motivation of chosing the canonical ensemble in Section~\ref{sec:sim-det}.}
 \section{Final remarks}
 

In the present work, we have shown that tetravalent particles with patches
located at the corners of a square belonging to the plane passing through
the particle center assembled into the cubic tetrastack crystals both in the bulk
and in the presence of the external field. However, the external field is crucial as it facilitates the 
formation of well-defined clusters with significantly fewer defects 
compared to the bulk phase. As the strength of the external field increases,
the epitaxial growth mechanism changes from a thin-thick film 
to a series of layering transitions as the adhesion increases, 
and has also an impact on whether the crystal will exhibit
a rough or smooth interface with the coexisting fluid. 


The selective formation of an open lattice such as cubic tetrastack
is of paramount importance as this polymorph exhibits a complete
photonic bandgap, making it potentially useful in terms of photonic crystal applications. 
Additionally, the crystalline structures remain stable when the external field is turned off and even upon heating, allowing one to test them for their photonic applications 
and also specific post-synthetic adjustments can be made. 

Although a selective formation of cubic tetrastack crystals has been reported for
several several systems involving star-polymers \cite{A6}
or triblock particles with triangular patches \cite{3d-tri-3}, 
the use of tetravalent particles with circular patches is arguably 
more feasible in terms of experimental realization. 
{\color{black} In particular, the latter is supported by the recent study
where the formation of tetrastack lattice has been predicted to be formed
by a four-component mixture of icosahedral and octahedral patchy particles \cite{liu2023}.
The authors confirmed its emergence using patchy DNA origami
and were able for the first-ever experimental formation of such an open lattice. 
Moreover, another experimental study demonstrated that }
random aggregates have been observed during the self-assembly of 
particles with triangular patches. The main
difficulty with directing the formation of well ordered structures is connected
with the lack of appropriate methods to obtain particles with triangular patches of appropriate size \cite{triangular-pine}. 
{\color{black} Therefore, we believe that another approach, using a different
geometry of patchy particles and most importantly in a one-component system, 
can be used in the near future for experimental
preparation of tetrastack crystals.}

\section*{Acknowledgments}
This work was supported by the National Science Centre, Poland, under Grant No. 2021/41/N/ST4/00437, PRELUDIUM 20.

\section*{Data Availability Statement}

The data that support the findings of this study are available from the corresponding author upon reasonable request.

\bibliography{biblio}

\begin{thebibliography}{10}

\bibitem{CR0}
I.~Sanchez-Burgos, E.~Sanz, C.~Vega, and J.~R. Espinosa, ``Fcc vs. hcp
  competition in colloidal hard-sphere nucleation: on their relative
  stability{,} interfacial free energy and nucleation rate,'' {\em Phys. Chem.
  Chem. Phys.}, vol.~23, pp.~19611--19626, 2021.

\bibitem{CR1}
A.~Mori, ``Computer simulations of crystal growth using a hard-sphere model,''
  {\em Crystals}, vol.~7, no.~4, 2017.

\bibitem{A1}
N.~R. Jana, L.~Gearheart, and C.~J. Murphy, ``Wet chemical synthesis of high
  aspect ratio cylindrical gold nanorods,'' {\em The Journal of Physical
  Chemistry B}, vol.~105, no.~19, pp.~4065--4067, 2001.

\bibitem{A2}
Y.~Sun and Y.~Xia, ``Shape-controlled synthesis of gold and silver
  nanoparticles,'' {\em Science}, vol.~298, no.~5601, pp.~2176--2179, 2002.

\bibitem{A3}
R.~Jin, Y.~Cao, C.~A. Mirkin, K.~L. Kelly, G.~C. Schatz, and J.~G. Zheng,
  ``Photoinduced conversion of silver nanospheres to nanoprisms,'' {\em
  Science}, vol.~294, no.~5548, pp.~1901--1903, 2001.

\bibitem{A4}
J.~P. Rolland, B.~W. Maynor, L.~E. Euliss, A.~E. Exner, G.~M. Denison, and
  J.~M. DeSimone, ``Direct fabrication and harvesting of monodisperse,
  shape-specific nanobiomaterials,'' {\em Journal of the American Chemical
  Society}, vol.~127, no.~28, pp.~10096--10100, 2005.
\newblock PMID: 16011375.

\bibitem{A5}
Y.~Yin, Y.~Lu, B.~Gates, and Y.~Xia, ``Template-assisted self-assembly: A
  practical route to complex aggregates of monodispersed colloids with
  well-defined sizes, shapes, and structures,'' {\em Journal of the American
  Chemical Society}, vol.~123, no.~36, pp.~8718--8729, 2001.
\newblock PMID: 11535076.

\bibitem{A8}
Y.-S. Cho, G.-R. Yi, J.-M. Lim, S.-H. Kim, V.~N. Manoharan, D.~J. Pine, and
  S.-M. Yang, ``Self-organization of bidisperse colloids in water droplets,''
  {\em Journal of the American Chemical Society}, vol.~127, no.~45,
  pp.~15968--15975, 2005.
\newblock PMID: 16277541.

\bibitem{A7}
S.~C. Glotzer and M.~J. Solomon, ``Anisotropy of building blocks and their
  assembly into complex structures,'' {\em Nature Materials}, vol.~6,
  pp.~557--562, Aug 2007.

\bibitem{jan1}
P.-G. de~Gennes, ``Soft matter (nobel lecture),'' {\em Angewandte Chemie
  International Edition in English}, vol.~31, no.~7, pp.~842--845, 1992.

\bibitem{jan2}
J.~Zhang, B.~A. Grzybowski, and S.~Granick, ``Janus particle synthesis,
  assembly, and application,'' {\em Langmuir}, vol.~33, no.~28, pp.~6964--6977,
  2017.
\newblock PMID: 28678499.

\bibitem{S2}
Z.~Zhang and S.~C. Glotzer, ``Self-assembly of patchy particles,'' {\em Nano
  Letters}, vol.~4, no.~8, pp.~1407--1413, 2004.
\newblock PMID: 29048902.

\bibitem{S3}
E.~Bianchi, R.~Blaak, and C.~N. Likos, ``Patchy colloids: state of the art and
  perspectives,'' {\em Phys. Chem. Chem. Phys.}, vol.~13, pp.~6397--6410, 2011.

\bibitem{jan3}
Y.~Iwashita and Y.~Kimura, ``Orientational order of one-patch colloidal
  particles in two dimensions,'' {\em Soft Matter}, vol.~10, pp.~7170--7181,
  2014.

\bibitem{size1}
H.~Shin and K.~S. Schweizer, ``Theory of two-dimensional self-assembly of janus
  colloids: crystallization and orientational ordering,'' {\em Soft Matter},
  vol.~10, pp.~262--274, 2014.

\bibitem{jan5}
S.~Jiang, J.~Yan, J.~K. Whitmer, S.~M. Anthony, E.~Luijten, and S.~Granick,
  ``Orientationally glassy crystals of janus spheres,'' {\em Phys. Rev. Lett.},
  vol.~112, p.~218301, May 2014.

\bibitem{jan7}
A.~G. Vanakaras, ``Self-organization and pattern formation of janus particles
  in two dimensions by computer simulations,'' {\em Langmuir}, vol.~22, no.~1,
  pp.~88--93, 2006.
\newblock PMID: 16378405.

\bibitem{kagome}
Q.~Chen, S.~C. Bae, and S.~Granick, ``Directed self-assembly of a colloidal
  kagome lattice,'' {\em Nature}, vol.~469, pp.~381--384, Jan 2011.

\bibitem{ord1}
F.~Sciortino, A.~Giacometti, and G.~Pastore, ``Phase diagram of janus
  particles,'' {\em Phys. Rev. Lett.}, vol.~103, p.~237801, Nov 2009.

\bibitem{ord2}
Z.~Preisler, T.~Vissers, F.~Smallenburg, G.~Munaò, and F.~Sciortino, ``Phase
  diagram of one-patch colloids forming tubes and lamellae,'' {\em The Journal
  of Physical Chemistry B}, vol.~117, no.~32, pp.~9540--9547, 2013.
\newblock PMID: 23902159.

\bibitem{ord3}
T.~Vissers, Z.~Preisler, F.~Smallenburg, M.~Dijkstra, and F.~Sciortino,
  ``{Predicting crystals of Janus colloids},'' {\em The Journal of Chemical
  Physics}, vol.~138, p.~164505, 04 2013.

\bibitem{S4}
F.~Romano, E.~Sanz, P.~Tartaglia, and F.~Sciortino, ``Phase diagram of
  trivalent and pentavalent patchy particles,'' {\em Journal of Physics:
  Condensed Matter}, vol.~24, p.~064113, jan 2012.

\bibitem{Kag1}
H.~Eslami, K.~Bahri, and F.~Müller-Plathe, ``Solid–liquid and solid–solid
  phase diagrams of self-assembled triblock janus nanoparticles from
  solution,'' {\em The Journal of Physical Chemistry C}, vol.~122, no.~16,
  pp.~9235--9244, 2018.

\bibitem{3d-tri-1}
Z.-W. Li, Y.-W. Sun, Y.-H. Wang, Y.-L. Zhu, Z.-Y. Lu, and Z.-Y. Sun,
  ``Softness-enhanced self-assembly of pyrochlore- and perovskite-like
  colloidal photonic crystals from triblock janus particles,'' {\em The Journal
  of Physical Chemistry Letters}, vol.~12, no.~30, pp.~7159--7165, 2021.
\newblock PMID: 34297560.

\bibitem{3d-tri-2}
Q.~Chen, E.~Diesel, J.~K. Whitmer, S.~C. Bae, E.~Luijten, and S.~Granick,
  ``Triblock colloids for directed self-assembly,'' {\em Journal of the
  American Chemical Society}, vol.~133, no.~20, pp.~7725--7727, 2011.
\newblock PMID: 21513357.

\bibitem{3d-tri-3}
F.~Romano and F.~Sciortino, ``Patterning symmetry in the rational design of
  colloidal crystals,'' {\em Nature Communications}, vol.~3, p.~975, Jul 2012.

\bibitem{3d-tri-4}
A.~B. Rao, J.~Shaw, A.~Neophytou, D.~Morphew, F.~Sciortino, R.~L. Johnston, and
  D.~Chakrabarti, ``Leveraging hierarchical self-assembly pathways for
  realizing colloidal photonic crystals,'' {\em ACS Nano}, vol.~14, no.~5,
  pp.~5348--5359, 2020.
\newblock PMID: 32374160.

\bibitem{ph1}
F.~Romano, E.~Sanz, and F.~Sciortino, ``Phase diagram of a tetrahedral patchy
  particle model for different interaction ranges,'' {\em The Journal of
  Chemical Physics}, vol.~132, no.~18, p.~184501, 2010.

\bibitem{ph2}
E.~G. Noya, C.~Vega, J.~P.~K. Doye, and A.~A. Louis, ``{The stability of a
  crystal with diamond structure for patchy particles with tetrahedral
  symmetry},'' {\em The Journal of Chemical Physics}, vol.~132, p.~234511, 06
  2010.

\bibitem{ph3}
Z.~Zhang, A.~S. Keys, T.~Chen, and S.~C. Glotzer, ``Self-assembly of patchy
  particles into diamond structures through molecular mimicry,'' {\em
  Langmuir}, vol.~21, no.~25, pp.~11547--11551, 2005.
\newblock PMID: 16316077.

\bibitem{light}
E.~Yablonovitch, ``Photonic band-gap structures,'' {\em J. Opt. Soc. Am. B},
  vol.~10, pp.~283--295, Feb 1993.

\bibitem{diam1}
D.~R. Nelson, ``Toward a tetravalent chemistry of colloids,'' {\em Nano
  Letters}, vol.~2, no.~10, pp.~1125--1129, 2002.

\bibitem{diam2}
M.~He, J.~P. Gales, {\'E}.~Ducrot, Z.~Gong, G.-R. Yi, S.~Sacanna, and D.~J.
  Pine, ``Colloidal diamond,'' {\em Nature}, vol.~585, pp.~524--529, Sep 2020.

\bibitem{tetrastack1}
T.~T. Ngo, C.~M. Liddell, M.~Ghebrebrhan, and J.~D. Joannopoulos, ``Tetrastack:
  Colloidal diamond-inspired structure with omnidirectional photonic band gap
  for low refractive index contrast,'' {\em Applied Physics Letters}, vol.~88,
  no.~24, p.~241920, 2006.

\bibitem{A6}
N.~A. Mahynski, L.~Rovigatti, C.~N. Likos, and A.~Z. Panagiotopoulos,
  ``Bottom-up colloidal crystal assembly with a twist,'' {\em ACS Nano},
  vol.~10, no.~5, pp.~5459--5467, 2016.
\newblock PMID: 27124487.

\bibitem{pyr1}
H.~Pattabhiraman, G.~Avvisati, and M.~Dijkstra, ``Novel pyrochlorelike crystal
  with a photonic band gap self-assembled using colloids with a simple
  interaction potential,'' {\em Phys. Rev. Lett.}, vol.~119, p.~157401, Oct
  2017.

\bibitem{pyr2}
\'{E}tienne Ducrot, J.~Gales, G.-R. Yi, and D.~J. Pine, ``Pyrochlore lattice,
  self-assembly and photonic band gap optimizations,'' {\em Opt. Express},
  vol.~26, pp.~30052--30060, Nov 2018.

\bibitem{liu2023}
H.~Liu, M.~Matthies, J.~Russo, L.~Rovigatti, R.~P. Narayanan, T.~Diep,
  D.~McKeen, O.~Gang, N.~Stephanopoulos, F.~Sciortino, H.~Yan, F.~Romano, and
  P.~Šulc, ``Inverse design of a pyrochlore lattice of dna origami through
  model-driven experiments,'' 2023.

\bibitem{ent1}
F.~Romano and F.~Sciortino, ``Two dimensional assembly of triblock janus
  particles into crystal phases in the two bond per patch limit,'' {\em Soft
  Matter}, vol.~7, pp.~5799--5804, 2011.

\bibitem{ent2}
X.~Mao, Q.~Chen, and S.~Granick, ``Entropy favours open colloidal lattices,''
  {\em Nature Materials}, vol.~12, pp.~217--222, Mar 2013.

\bibitem{tau}
M.~Trau, D.~A. Saville, and I.~A. Aksay, ``Field-induced layering of colloidal
  crystals,'' {\em Science}, vol.~272, no.~5262, pp.~706--709, 1996.

\bibitem{temp1}
A.~D. Dinsmore, A.~G. Yodh, and D.~J. Pine, ``Entropic control of particle
  motion using passive surface microstructures,'' {\em Nature}, vol.~383,
  pp.~239--242, Sep 1996.

\bibitem{temp2}
A.~van Blaaderen, R.~Ruel, and P.~Wiltzius, ``Template-directed colloidal
  crystallization,'' {\em Nature}, vol.~385, pp.~321--324, Jan 1997.

\bibitem{temp3}
W.~Lee, A.~Chan, M.~A. Bevan, J.~A. Lewis, and P.~V. Braun,
  ``Nanoparticle-mediated epitaxial assembly of colloidal crystals on patterned
  substrates,'' {\em Langmuir}, vol.~20, no.~13, pp.~5262--5270, 2004.
\newblock PMID: 15986661.

\bibitem{MC-temp}
A.-P. Hynninen, J.~H.~J. Thijssen, E.~C.~M. Vermolen, M.~Dijkstra, and A.~van
  Blaaderen, ``Self-assembly route for photonic crystals with a bandgap in the
  visible region,'' {\em Nature Materials}, vol.~6, pp.~202--205, Mar 2007.

\bibitem{LB1}
{\L}.~Baran, D.~Tarasewicz, D.~M. Kami{\'n}ski, and W.~R{\.z}ysko, ``Pursuing
  colloidal diamonds,'' {\em Nanoscale}, vol.~15, pp.~10623--10633, 2023.

\bibitem{Grzybowski}
K.~J.~M. Bishop, C.~E. Wilmer, S.~Soh, and B.~A. Grzybowski, ``Nanoscale forces
  and their uses in self-assembly,'' {\em Small}, vol.~5, no.~14,
  pp.~1600--1630, 2009.

\bibitem{steinhardt}
P.~J. Steinhardt, D.~R. Nelson, and M.~Ronchetti, ``Bond-orientational order in
  liquids and glasses,'' {\em Phys. Rev. B}, vol.~28, pp.~784--805, Jul 1983.

\bibitem{Kag2}
H.~Eslami, P.~Sedaghat, and F.~Müller-Plathe, ``Local bond order parameters
  for accurate determination of crystal structures in two and three
  dimensions,'' {\em Phys. Chem. Chem. Phys.}, vol.~20, pp.~27059--27068, 2018.

\bibitem{LAMMPS}
A.~P. Thompson, H.~M. Aktulga, R.~Berger, D.~S. Bolintineanu, W.~M. Brown,
  P.~S. Crozier, P.~J. in~'t Veld, A.~Kohlmeyer, S.~G. Moore, T.~D. Nguyen,
  R.~Shan, M.~J. Stevens, J.~Tranchida, C.~Trott, and S.~J. Plimpton,
  ``{LAMMPS} - a flexible simulation tool for particle-based materials modeling
  at the atomic, meso, and continuum scales,'' {\em Comp. Phys. Comm.},
  vol.~271, p.~108171, 2022.

\bibitem{nhchains}
G.~J. Martyna, M.~L. Klein, and M.~Tuckerman, ``{Nosé–Hoover chains: The
  canonical ensemble via continuous dynamics},'' {\em The Journal of Chemical
  Physics}, vol.~97, pp.~2635--2643, 08 1992.

\bibitem{AD1}
R.~Pandit, M.~Schick, and M.~Wortis, ``Systematics of multilayer adsorption
  phenomena on attractive substrates,'' {\em Phys. Rev. B}, vol.~26,
  pp.~5112--5140, Nov 1982.

\bibitem{free2}
C.~Vega and E.~G. Noya, ``{Revisiting the Frenkel-Ladd method to compute the
  free energy of solids: The Einstein molecule approach},'' {\em The Journal of
  Chemical Physics}, vol.~127, p.~154113, 10 2007.

\bibitem{triangular-pine}
M.~He, J.~P. Gales, X.~Shen, M.~J. Kim, and D.~J. Pine, ``Colloidal particles
  with triangular patches,'' {\em Langmuir}, vol.~37, no.~23, pp.~7246--7253,
  2021.
\newblock PMID: 34081481.

\end{thebibliography}
\bibliographystyle{ieeetr}

\end{document}